\documentclass[12pt]{article}

\input{epsf}
\usepackage[dvips]{graphicx}
\usepackage{cite}
\def\1ad{\mbox{\normalsize $^1$}}
\def\2ad{\mbox{\normalsize $^2$}}
\def\3ad{\mbox{\normalsize $^3$}}
\def\4ad{\mbox{\normalsize $^4$}}
\def\5ad{\mbox{\normalsize $^5$}}
\def\6ad{\mbox{\normalsize $^6$}}
\def\7ad{\mbox{\normalsize $^7$}}
\def\8ad{\mbox{\normalsize $^8$}}

           \def\CO{{\cal O}} 
 \def\CC{{\cal C}}

\def\gev{\mbox{ GeV}}

\def\threedx{\mbox{$(3+\delta x)$}} 
\def\dj{\hbox{d\kern-0.347em \vrule width 0.3em height 1.252ex depth
-1.21ex \kern 0.051em}}

\def\Re{{\rm Re\,}}

\def\ket{\rangle}
\def\bra{\langle}

\def\balpha{\overline \alpha}

\def\bS{\overline S}
\def\bM{\overline M}
\def\bT{\overline T}

\def\bW{\overline W}
\def\bphi{\overline \phi}
\def\balpha{\overline \alpha}
\def\bbeta{\overline \beta}

\def\pt{\partial}

\newcommand{\tev}{\mbox{TeV}}
\newcommand{\hc}{\mbox{h.c.}}
\newcommand{\be}{\begin{equation}}
\newcommand{\ee}{\end{equation}}
\newcommand{\ben}{\begin{equation*}}
\newcommand{\een}{\end{equation*}}
\newcommand{\ba}{\begin{eqnarray}}
\newcommand{\ea}{\end{eqnarray}}
\newcommand{\ban}{\begin{eqnarray*}}
\newcommand{\ean}{\end{eqnarray*}}
\newcommand{\brr}{\begin{array}}
\newcommand{\err}{\end{array}}
\newcommand{\bc}{\begin{center}}
\newcommand{\ec}{\end{center}}

\newcommand{\bea}{\begin{eqnarray}}
\newcommand{\eea}{\end{eqnarray}}
\newcommand{\bean}{\begin{eqnarray*}}
\newcommand{\eean}{\end{eqnarray*}}

\newcommand{\nn}{\nonumber }

\newcommand{\eg}{\mbox{\it e.g.~}}
\newcommand{\ie}{\mbox{\it i.e.~}}
\newcommand{\vev}[1]{\mbox{$\langle #1 \rangle $}}
\newcommand{\leqsim}{\,\raisebox{-0.6ex}{$\buildrel < \over \sim$}\,}
\newcommand\lsim{\mathrel{\rlap{\lower4pt\hbox{\hskip1pt$\sim$}}
    \raise1pt\hbox{$<$}}}
\newcommand\gsim{\mathrel{\rlap{\lower4pt\hbox{\hskip1pt$\sim$}}
    \raise1pt\hbox{$>$}}}
\begin{document} 

\begin{titlepage}

\title{\bf {CP and Flavour \\
in Effective Type I String Models}}

\author{ S.A. Abel $^{a\, b}$ and G. Servant $^{c}$ }

\maketitle

\vskip 15pt

\centerline{$^{a}$ {\it CPES, University of Sussex, Falmer, Brighton
BN1 9RH, UK}}
\vskip 3pt
\centerline{$^{b}$ {\it IPPP, 
Durham University, South Road, Durham DH1 3LE, UK}}
\vskip 3pt
\centerline{$^{c}$ {\it CEA-SACLAY, SPhT,
F-91191 Gif-sur-Yvette C\'edex, France.}}

\vglue .5truecm

\begin{abstract}
\vskip 3pt

\noindent
Effective type I string models allow 
stabilization of the dilaton and moduli 
fields with only a single gaugino 
condensate. We show that, as well as breaking supersymmetry, 
the stabilization can spontaneously break CP.
We find that this source of CP violation 
hints strongly at a natural solution to the supersymmetric 
CP and flavour problems. Even though the CP violation generates
physical phases in the Yukawa couplings, all the supersymmetry breaking
terms are found to be automatically real and given by the 
$U(1)$ charges of the associated
Yukawa couplings.  These can be chosen to 
have a structure (degenerate or non-universal) which
suppresses FCNCs and EDMs. We examine the phenomenological implications, 
including the generation of the $\mu$-term, and the effect 
of higher order terms.
\end{abstract}

\begin{flushright}
Saclay t00/179\
\end{flushright}

\end{titlepage}

\section{Introduction}

Flavour and CP are especially 
problematic in supersymmetry because a {\em generic} choice of
parameters violates experimental bounds on, for instance, 
$b\rightarrow s \gamma $ and neutron electric dipole moments (EDMs).
These two aspects of supersymmetry are known generically as the SUSY 
flavour and CP problems, and they are probably the most
useful tools for probing the underlying theory.
 In this paper we present a dynamical solution 
to these problems, which emerges 
in the light of recent progress on dilaton stabilization
in effective models of type I string~\cite{as}. 


Since the importance of dilaton stabilization may be 
less than transparent, let us begin by discussing 
the canonical example of a `dynamical' solution
to the SUSY flavour and CP problems, 
dilaton domination. The idea  is illustrated 
schematically in fig.(1a).
Supersymmetry breaking is described by the {\it vevs} of the auxilliary ($F$)
fields. Together they describe a vector whose length 
is determined by the requirement that the cosmological constant 
be zero. Its direction however is determined by whatever dynamics 
breaks supersymmetry. 
Dilaton domination asserts that it is aligned with the dilaton.
Since the dilaton couples equally to all fields, 
the resulting SUSY breaking terms in the visible sector are 
very constrained and indeed one finds a suppression of EDMs and FCNCs. 

The assumptions underlying dilaton domination are rather
more brutal than they might at first appear,
since there are more fields than just the dilaton and 
moduli involved in Planck scale physics. For example, 
any superfield whose scalar component gets a {\it vev} at a high scale can 
also be involved in transmitting supersymmetry breaking.
In particular this is likely to be the case for the very fields that are 
responsible for flavour structure and CP violation in the 
first place. Thus one has to assume that, 
either the spontaneous breaking of CP and flavour does not contribute significantly 
to supersymmetry breaking thereby affecting the dynamics (\eg the goldstino 
angle in the case of dilaton domination), or
that the alignment of the goldstino with the dilaton 
is true {\em a posteriori}.

Clearly, a credible `dynamical' solution requires 
a full determination of the goldstino direction, and 
that in turn requires a specific model of dilaton and 
moduli stabilization and spontaneous CP violation.
Without all of these ingredients, we 
think that any dynamical solutions to the SUSY 
flavour and CP problems will be at best incomplete.
To put it more bluntly; can one really trust 
a dynamical solution to the SUSY CP problem that does not explain 
the origin of CP violation?

These considerations suggest the approach that we will follow in this
paper, which is to avoid tackling flavour and CP head on,
but rather to begin by attacking the most difficult part of the 
problem, namely dilaton stabilization. Our starting point will be the dilaton
stabilization scheme found in ref.\cite{as} in effective type I
models. As we shall see, this scheme includes a rather generic way to
spontaneously break CP.  This gives us the required complete dynamical
picture of dilaton and moduli {\it vevs}, supersymmetry breaking {\em and}
CP breaking. 

\begin{figure}[t*]
\begin{center}
\hspace*{-0.6in}
\epsfxsize=5in
\epsffile{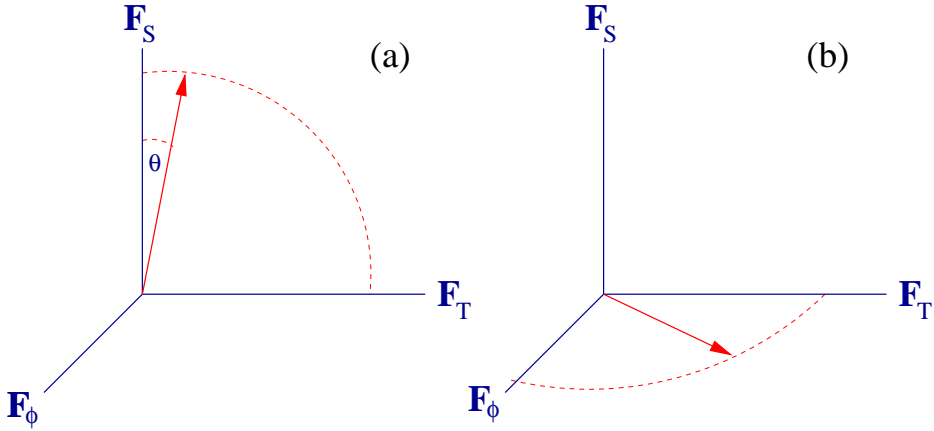}
\vspace{0in}
\caption{\it Supersymmetry breaking dynamics leads to non-zero {\it vevs} 
of auxilliary fields of the dilaton, moduli, and any other fields, 
such as Froggatt-Nielsen fields, that get {\it vevs} at high scales. Their 
relative sizes can be expressed with `goldstino' angles.
(a) shows schematically the dilaton domination scenario. (b) 
shows effective  type I models with supersymmetry broken by a condensing 
D9-brane gauge group, and with stabilized dilaton and moduli fields. 
The full dynamical calculation relates the $F$-terms and 
completely determines the goldstino angles.}
\label{v0}
\end{center}
\end{figure}

Anomalous $U(1)$'s play a central role in the 
stabilization, and consequently the supersymmetry breaking picture which 
emerges bears some resemblance to the 
anomalous $U(1)$ mediation models of refs.\cite{anomed},
although the $F$-terms are the dominant source of supersymmetry 
breaking in the case we examine, rather than the $D$ terms. 
The picture of supersymmetry breaking that eventually emerges is 
shown in fig.(1b). The dilaton auxilliary field is zero 
so supersymmetry breaking is {\em not} dilaton dominated. Nevertheless 
the goldstino direction is determined 
by the choice of $U(1)$ charges.
An appropriate choice
gives soft terms that are degenerate providing a leading order suppression of 
FCNCs that depends only on the charge assignments, and 
is otherwise independent of the form of Yukawa couplings.
In addition, independently of the charge assignments, the soft terms are guaranteed to be real
even though the Yukawa couplings can have maximal CP violation, thereby
suppressing EDMs.
There can, however, be a higher order 
parametrically small breaking of CP in the soft terms as
well. If the CP phase in the CKM matrix happens to be 
small, this is an explicit manifestation of the approximate CP idea \cite{nir}.

We stress that we will not make any 
ad-hoc assumptions about the dilaton or moduli stabilization, 
or the CP violation which we 
will treat completely. Moreover, the suppression of FCNCs and EDMs is 
extremely general, requires no assumptions about the hidden sector 
particle content 
and only very mild assumptions about the hidden sector superpotential.
Consequently the SUSY breaking in the visible sector can be much more general 
than the dilaton domination pattern. 

We begin in the following section by 
recapitulating the supergravity scalar potential obtained in 
ref.\cite{as} and showing how it stabilizes the dilaton and 
moduli. Here we will discuss the role of modular invariance in the 
non-perturbative superpotential, and also demonstrate why a similar 
stabilization mechanism cannot work in the heterotic string.
In section III, we present our model for generating Yukawa hierarchies and 
 see how CP can be spontaneously broken. In section IV, we
 discuss the condensation and string
scales allowed by our model. 
Section V is devoted to the computation of the SUSY 
breaking terms and we
show how CP violation is naturally suppressed in the latter  
even though it appears in the Yukawa couplings.
In section VI we address the problem of the generation of the $\mu$-term and the 
implications of higher order corrections for the susy flavour and CP problems.
We will find that only a mild tuning of couplings 
can potentially solves these problems.
We summarize our results in section VII.

\section{Stabilization with and without modular invariance}

In this section we discuss the main features of the dilaton stabilization 
mechanism derived in ref.\cite{as} in the context of effective type IIB 
orientifold models. 
We first introduce the effective models, discuss the role of modular invariance
in the effective potential, and state the two main assumptions that lead to 
a stabilization of the dilaton and moduli.
We then consider theories both with and without modular invariance.
The first case was discussed in ref.\cite{as}, and we shall 
briefly recap the results and generalize them.
We then consider models in which the superpotential
is constrained by modular symmetries. The modular invariant case has the 
advantage that the K\"ahler potential can be adjusted to 
give a vanishing cosmological constant at the (local) minimum. 
We also discuss why a similar minimization 
cannot be achieved in the heterotic string.  

\subsection{ Effective models and modular invariance}

\noindent Our starting point is the effective theory of 
$D=4$, $N=1$ type IIB orientifolds. The important features are 
as follows:

As well as the matter fields, the models contain  
a complex dilaton $S$, untwisted moduli $T_i$ 
associated with the size and shape of the extra dimensions 
and complex superfields
$M_k$ associated with the fixed points (labelled by $k$) of the 
underlying orbifold. An important property of these models 
is that the $M_k$ superfields appear
linearly in the gauge kinetic functions.
For gauge groups living on a D9-brane
\be
f_{9\, a} = S+\sum_k \sigma^k_a M_k \, ,
\ee
whereas for the D5-branes
\be
f_{5_i \, a}= T_i+\sum_k \sigma_{ia}^k M_k\, ,
\ee
where $\sigma_a^k$ are calculable model dependent coefficients and
$k$ runs over the different twisted sectors. 
In most of what follows we will consider only one degenerate 
value for the $M_k$ superfields 
which we will denote $M$ (it is straightforward to generalize).
The $M_k$ superfields participate in the generalized Green--Schwarz
mechanism for the cancellation of $U(1)_X$ anomalies \cite{sagnotti}. This 
contrasts 
with heterotic models where the dilaton plays this role. Under a $U(1)$
transformation through a phase $\alpha$, 
the $M_k$ fields transform linearly, 
\be 
M_k \rightarrow M_k+i\frac{\delta_{GS}}{2}\alpha\, .
\ee
Type IIB orientifold / heterotic duality has been used to
argue that there is also a $\sigma$-model invariance under 
$SL(2,{\bf Z})$ transformations of the $T_i$;
\be 
T_i \rightarrow \frac{a_i T_i - i b_i}{i c_i T_i + d_i} 
\hspace{0.3cm};\hspace{0.5cm} \phi_n \rightarrow \phi_n \prod_{i=1}^3 (i c_i T_i + d_i)^{n_n^i} \, 
\ee
where $a_i,~b_i,~c_i,~d_i\in {\cal Z}$ and $n_n^i$ are modular weights of the 
$\phi_n$ with respect to the ith complex direction. These symmetries 
are broken by the presence of D5-branes, as is obvious from the 
expressions for $f_{5i}$. However, one expects a
remnant of them to survive in directions that are orthogonal to the D5 branes, 
and again it is the $M_k$ fields that 
shift to cancel any $\sigma$-gauge anomalies~\cite{iru}\footnote{For further work 
on the cancellation of $SL(2,{\bf Z})$/gauge/gravitational anomalies in orientifold 
models at string level see \cite{Scrucca}.};
\be 
M_k \rightarrow M_k + \sum_i \delta^i_k \ln ({i c_i T_i + d_i})\, .
\ee
In order to cancel $\sigma-F_a$ anomalies, denoted $C_a^i$, we require 
\be 
\sum_k \sigma^k_{ia} \delta^i_k = C_a^i \, ,
\ee
for any preserved modular symmetries.
The anomalous $U(1)$'s and modular symmetries will be important 
constraints on the possible form of the superpotential.

When taking into account the presence of $D5$ branes in the vacuum,
there are four types of charged matter fields:
$C_i^9$ ($i$ labels the three complex dimensions) comes from open strings 
starting and ending on the 9-branes; 
$C_i^{5_j}$ from open strings starting and ending on the same $5_i$-branes;
$C^{5_i5_j}$ from open strings starting and ending on different sets of 
$5_i$-branes; $C^{95_i}$ from open strings with one end on the 9-branes 
and the other end on the $5_i$-branes. The K\"ahler potential for the $S$,
 $T_i$ and $C$ fields 
is of the general form~\cite{imr}:
\ba 
\label{completeK}
K &=& -\ln \left( S+\bS - \sum_i |C_i^{5_i}|^2  \right) - \sum_i 
\ln\left( T_i+\bT_i
 - |C_i^9|^2 
-  \sum_{j\neq k\neq i = 1}^3 |C_j^{5_k}|^2 \right) \nn\\
&& + \frac{1}{2}\sum_{j\neq k\neq i = 1}^3 \left(
\frac{|C^{5_j\, 5_k}|^2}{(S+\bS)^{1/2} (T_i+\bT_i)^{1/2} }
+ \frac{|C^{95_i}|^2}{(T_j+\bT_j)^{1/2}  (T_k+\bT_k)^{1/2}} \right)\, .
\ea
We will only consider $C_i^9$ because they correspond to the fields which will later condense. 
$K$ can be rewritten at one loop:
\be
\label{Ka}
K = - \ln\, s - 3 \ln \tau + \hat{K}(m),  
\ee
where 
\be 
s = S + \bS \, ;\ \ \, \tau = T + \bT  - \sum_n|\phi_n|^2
\,;\ \ \, m = M + \bM - \delta \ln \tau\, .
\ee
We have introduced generic fields $\phi_n$ to represent 
matter fields $C_i^9$ and for the moment consider only the overall moduli, 
taking $T_i = T$, $M_k=M$ and $\delta^i_k=\delta$ leaving the more general 
case for later.
We shall express all quantities including the string scale ($M_s$)
in natural units where $M_P=1$ and for later convenience define fields
scaled in string units with a tilde -- for example $\tilde{\phi} = \phi/M_s$.
The first two terms have the usual ``no-scale'' structure
with the $T$ and $\phi_n$-dependence appearing in the combination $\tau$ only.
All $M$ dependence appears in the modular invariant combination, $m$,
and giving a {\it vev} to $m$ takes us away from the orientifold point.
The $\delta \ln \tau$ correction
can be deduced from the one loop expression for the gauge 
coupling and depends on the tree-level expression for $K$.
Although it is currently unclear what the precise form of the 
$m$-dependence in the K\"ahler potential should be, we know that 
$\hat{K}$ is an even function of 
$m$ thanks to the orbifold symmetry, and that 
the leading term in an expansion about the orientifold point,
$m=0$, is quadratic, $\hat{K}=  \frac{1}{2} m^2 + \ldots $. 
We will accommodate the uncertainty in the form of $\hat{K}$ 
by working with the parameter 
$x=\partial \hat{K}/\partial M $ where near the orientifold point 
$x\approx m$.

\subsection{ Two assumptions for stabilization}

\noindent This completes the general overview of the type I models 
that will form the basis for our discussion\footnote{For alternative 
studies on supersymmety breaking in type I strings see for 
example \cite{ads}.}. In 
order to stabilize the dilaton, we now augment them with two 
mild assumptions about the superpotential:

\begin{itemize}

\item Our first assumption is that there is a non-perturbative 
contribution to the superpotential, $W_{np}$, which 
is generated by hidden sector
gaugino condensation with single gauge group $SU(N_c)$ residing on a 
D9-brane and with 
extra (anti)quarks ($\overline{{\cal Q}}$) ${\cal Q}$ in the (anti)fundamental 
representation of $SU(N_c)$. Below the scale $\Lambda=M_se^{-f_9/2\beta}$, 
where $\beta= \frac{3N_c-1}{16\pi^2}$,
$\overline{{\cal Q}}$  and ${\cal Q}$ form a composite meson field,
$ \phi_0=\sqrt{{\cal Q} \overline{\cal Q}}$.
$W_{np}$ is fixed uniquely by global symmetries and reads
\be
\label{fix2aa}
W_{np} = \left( \frac{\Lambda ^{3 N_c-1}}{\phi_0^2}
\right)^{\frac{1}{N_c-1}}\, .
\ee
There is no $T$-dependence in this expression since
there is no $T$-dependence in the 
one-loop expression for the gauge kinetic function 
$f_9$ in the type I case. 
Note that the superpotential is also invariant under
the $U(1)_X$ symmetry, and has the correct modular behaviour.
For example, the lagrangian is invariant 
 under overall modular transformations if 
the combination $G=K+\ln |W|^2$ is invariant, which implies 
that $W_{np}$ has weight $-3$. In models where there are 
no D5-branes, this is the case and the necessary modular weight is 
provided by $\phi_0$ and the transformation of $M$;
$Z_3$ and $Z_7$ orientifolds
have $f_9 = S \pm \beta M$ with our definitions\footnote{Note that we are 
using the definition $Re(f)=1/g^2$, and the conventions of ref.\cite{beatriz} 
in which $\beta = (3N_c-1)/16\pi^2$.}, 
and under an $SL(2,{\cal Z})$ transformation
$M\rightarrow M \pm 2  \ln ({i c T + d})$~\cite{iru}. Thus both $\Lambda$ 
and $\phi_0$ have weight $-1$, and $W_{np}$ has overall weight $-3$ as 
required. This is also true when there are $N_m$ mesons, in which
case it is det$(\tilde{\cal Q}{\cal Q})$ (with weight $-2 N_m$) that appears 
in the denominator. As we mentioned above, the $T_i$ modular symmetries are 
broken by the presence of D$5_i$-branes, but are expected to be 
preserved along the directions without them.

\item
The second assumption is that, as well as the MSSM, the superpotential 
contains additional pieces involving the remaining 
fields $\phi_{m=1,..N}$. In particular the extra terms 
should generate a perturbative 
mass term for $\phi_0$ (\eg $\phi_1 \phi_0^2$).
The $\phi_m$ and $\phi_0$ are charged 
under the anomalous $U(1)_X$ with charges $q_{m,0}$, so that the 
perturbative piece of the hidden sector superpotential 
can always be written in terms of 
the $m$ invariants of $U(1)_X$ which we can choose arbitrarily to be
$X_m={\phi}_m^{- 2 q_0 / q_m} {\phi}_0^2$.

\end{itemize}

\subsection{The general form of the scalar potential}

\noindent In type I models, the K\"ahler metric can be inverted without approximation, 
and the resulting scalar potential takes the form~\cite{as}
\be 
V_F=e^G B \, ,
\label{vf}
\ee 
where
\ba
G&=&K+\ln|W|^2  \, \nn\\
B&=&\delta B_0 + \left|1-s\frac{W_S}{W}\right|^2 + 
\frac{\tau}{(3+\delta x)}\sum_n\left|\frac{W_n}{W}\right|^2 
+  
\hat{A}
\left| B_0+\sigma \frac{W_S}{W}\right|^2 \nn \\
B_0 &=& x - \frac{\delta}{\hat{A}} \nn \\
\hat{A} &=&  \frac{1}{ \hat{K}_{M\bM} }+ \frac{\delta^2}{3+\delta x}
\, ,
\eea
and where subscripts denote differentiation, and it is convenient to define
$x= K_M$. The potential can be expressed more concisely in terms of the auxilliary fields,
\be 
F^\alpha = e^{-G/2} G^{\alpha \overline{\beta}} G_{\overline{\beta}}\, .
\ee
The $F$-terms that we need are 
\ba 
F^S &=& e^{G/2} \, . s\left( s\frac{\bW_S}{\bW} - 1\right)\nn\\ 
F^M &=&  e^{G/2} \, . \hat{A}  
\left( B_0 + \sigma \frac{\bW_S}{\bW}\right) 
 \nn\\
F^0 &=& \phi_0 e^{G/2}  \, .\frac{1}{|\hat{\phi}_0|^2} 
\left( 2\sum _m \frac{X_m \bW_{X_m}}{\bW} 
+\frac{1}{4\pi^2} \frac{\bW_S}{\bW}\right) \nn\\
F^m &=& \phi_m e^{G/2}  \, .\frac{-2q_0}{q_m |\hat{\phi}_m|^2}
\frac{X_m \bW_{X_m}}{\bW}  \, .
\ea
Above and henceforth, we denote canonically normalized fields with a hat;
\be
\hat{\phi}_n = \sqrt{\frac{\threedx}{\tau} } \phi_n \,\, ; \,\,
\hat{F}^n = \sqrt{\frac{\threedx}{\tau} } F^n \,\, ; \,\,
\hat{F}^S= \frac{1}{s} F^S  \,\, ; \,\,
\hat{F}^M= \frac{F^M}{\sqrt{\hat{A}}} \, .
\ee
In terms of the canonically normalized auxilliary fields we have 
\be 
V_F = e^G \delta B_0 + |\hat{F}^S|^2  + |\hat{F}^M|^2 + 
\sum_{n = 0} |\hat{F}^n|^2 \, .
\ee
This form is reminiscent of the no-scale models, however 
the potential is not positive definite 
since the functions $B_0$ can be negative for finite values of $x$. 
Indeed, when $\hat{F}^{S,M,n}=0$, 
supersymmetry is restored where $3+\delta x = 0$. Here 
we find the global minimum with $V_F = -3 e^G$.

In addition to the $F$-term contribution, there is an important 
$D$-term contribution coming from the anomalous $U(1)$, which
takes the form
\be
V_D=\frac{g_X^2}{2} D_X^2\, ,
\ee
where
\be
D_X 
= q_n |\hat{\phi}_n|^2 - \delta _{GS} \frac{x}{2}\, .
\ee
The Fayet--Iliopoulos term,
\be 
\xi^2=\frac{1}{2}\delta_{GS}x \, ,
\ee
is given by $m$ as opposed to $S$ and so can be zero~\cite{iru,uranga}. 

Before tackling the stabilization in detail,
let us first highlight the general features of the potential that make 
it possible. 
The important aspect of the $V_F$ contribution to the potential is that $W_S/W$ appears 
{\em twice}, both in $\hat{F}^M$ and $\hat{F}^S$, due to the $M$ contribution to the 
gauge kinetic function. In previous work, the assumption has usually been that 
integrating out the mesons leads to $W\sim e^{-8\pi^2 S/\beta} $ so that $W_S/W$ 
inevitably ends up being a {\em negative} constant. In the present type I case however, we
leave $W_S/W$ as a fully dynamical variable and, by employing an additional perturbative contribution 
to the superpotential, set its {\it vev} to be {\em positive} thanks to the $F^M$ contribution. 
The dilaton then trivially finds its minimum where $\hat{F}^S=0$.
The $D$-term contribution is important because it forces a local minimum 
at a finite value of $x$, where supersymmetry is broken. There is thus an interesting 
interplay between the $F$ and $D$-term contributions.

\subsection{Stabilization without modular invariance}

\noindent In ref.\cite{as} it was shown that our two assumptions can lead to a natural
stabilization of the dilaton. In that example, which we shall now briefly recap, 
the superpotential is of the form 
\be
\label{toosimple}
W=W_{np}+W_p\, ,
\ee
where the perturbative part of the superpotential, $W_p(X_m)$, 
is some function of gauge invariants.

\begin{figure}[t*]
\begin{center}
\epsfxsize=5.8in
\epsffile{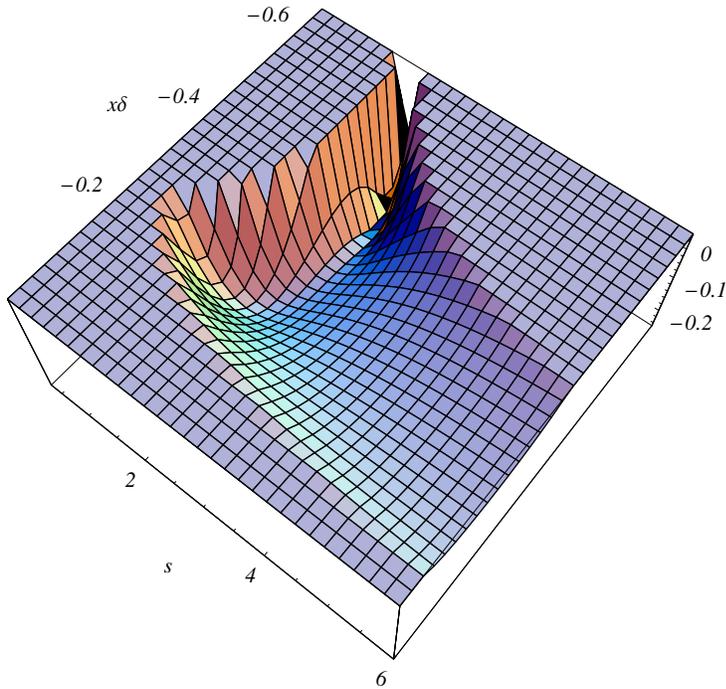}
\vspace{-3.2in}
\caption{\it The $F$-part of the potential 
(\ref{vf}) in the plane ($s,x$) where $s=S+\bS$ and 
$x=\frac{\partial{\hat K}}{\partial M}\sim m$ at small $m$. The 
valley approaches $x=0$ at large dilaton values. 
Thus, starting at the orientifold point
$m=0$, the field can roll down this valley from large $s$ values 
to its minimum.}
\label{v1}
\end{center}
\end{figure}

Let us first assume that 
$x=K_M$ has a non-zero value and determine the corresponding {\it vevs} of all the other fields. 
This will lead to a potential purely in $x$, whose minimization 
we shall consider at the end.
The $D$-term clearly dominates the potential in any 
reasonable model (with $e^G\sim m_W^2$), 
so we begin the minimization by as usual imposing $D_X = 0$;
\be
\label{dterm}
\sum_n q_n |\hat{\phi}_n|^2 =
\frac{|\delta_{GS}x|}{2} \, ,
\ee
where we have anticipated that 
$\langle \delta_{GS} x \rangle$ will eventually be positive.
This equation determines $\tau$ in terms of the other fields, 
since it does not appear elsewhere in the potential.
For definiteness we will take $q_0 > 0$.

The minimization of $V_F$ is equivalent to minimizing $B$ 
if the final cosmological constant is small or zero
as we must check at the end. The independent variables are 
$s$, $\phi_{m>0}$ and $\phi_0$, but  
things simplify greatly if we can trade them; 
$\phi_m$ for $X_m$, and $\phi_0$ for $W_S/W$. 
We can do this if, defining 
\be 
\rho_m = \frac{q_m |\phi_m|^2 }{ q_0 |\phi_0|^2} \, , 
\ee
the solutions satisfy $\rho_m\ll 1$. We shall see shortly how 
this can be achieved.

Minimization under this approximation gives simple but
non-trivial relations between the 
auxilliary fields;
\ba
\label{imp_relation0}
F^0 &=&
\frac{q_0 \phi_0}{q_m \phi_m} F^m = -4 \pi^2 \sigma \phi_0F^M =
 \frac{e^{G/2}}{s} \frac{q_0 \phi_0}{2\pi^2  |\delta_{GS}.x|}\nn\\
F^S &=&0 \, ,
\ea
to leading order in $\rho_m$.
These dynamical relations will be important in determining the 
behaviour of the soft terms. In particular they ensure nice properties
such as reality of $A$-terms.
Note that without $W_{np}$ the minimizations separate and 
we trivially find $F_n=0$. Thus supersymmetry is always unbroken 
before condensation.

The resulting expression for the stabilized dilaton is 
\be
s(x)  =  -\frac{\sigma^2 + \lambda^2}{\sigma B_0} + {\cal O}(\rho_m)\, ,
\ee
where we have defined 
\ba
\lambda^2 &=& \frac{1}{\hat{A}}.\frac{q_0}{8\pi^4
|\delta_{GS}x| } \, .
\ea 
The remaining equations, determining the {\it vevs} of 
$\phi_0$ and $\phi_m$ respectively, 
can be written in terms of $s(x)$;
\ba
\frac{W_S}{W}
& = & \frac{1}{s}  \nn\\
X_m \frac{W_{X_m} }{W}
& = & \frac{q_m \rho_m }{q_0 8 \pi^2 s} + {\cal O}(\rho_m^2)\, .
\label{solutions0}
\ea
The first requirement for this solution, and our assumption, 
to be consistent, is that the perturbative
superpotential has couplings such that 
$\partial W_p / \partial X_m \approx 0$ gives $\vev{W_p}\neq 0$, 
and the assumed $\rho_m \ll 1$.
The second requirement is of course that $s$ is positive
(and hence that $B_0(x)/\sigma$ is negative) for the particular 
value of $x$.

On examining the remaining $x$-dependent potential,
we find that a minimum can develop where all of these conditions are 
satisfied; assuming that we have chosen parameters 
such that $\rho_{m}\ll 1 $, 
the potential is given by 
\be 
\label{BB}
B = B_0 \frac{\delta }{\sigma^2+\lambda^2}
\left(
\sigma^2 - \sigma_0^2
\right) \, ,
\ee
where $\sigma_0^2$ is a constant; $\sigma_0^2 = \frac{q_0 }
{ 8\pi^4 \delta_{GS}\delta} $.
Two options can now be considered for determining the value of $x$;
\begin{itemize}

\item The no-scale option: Set the cosmological constant to be exactly 
zero. As we have seen, $\sigma B_0$ must be non-zero and negative to 
stabilize the 
dilaton. However we can make the potential completely flat 
by choosing $\sigma=\sigma_0$. 
The $x$ dependent {\it vevs} for $\phi_m$, $\phi_0$, $\tau$ and $s$ 
then correspond to a flat direction with 
zero cosmological constant. A natural possibility 
is that $x$ and therefore $M$ can be fixed 
by minimizing the potential after radiatively induced 
electroweak symmetry breaking.

\item The K\"ahler stabilization option: 
particular forms of $\hat{K}$ can  give a local minimum 
in the function $B_0(x) $ with a small but negative cosmological constant of order $-\delta^2$. 
The minimization condition is found to be 
\be
\label{mincond}
\delta \frac{\partial K_{M\bM}}{\pt x} = 1 + \frac{2 \delta^2 K_{M\bM}}{3+\delta x} \, .
\ee
For example when 
\be 
\hat{K}(m) = \frac{3}{a} \ln\left( 1+ \frac{a}{6} m^2 \right)\, ,
\ee
the minimum in $\delta B_0$ is at $\delta x<0$, and 
gives a cosmological constant   
\be 
\delta B_0|_{min} \approx \frac{1}{2 a} - \delta^2 .
\ee
In conjunction with the $D=0$ condition, this
implies that $\delta_{GS} \delta q_0 > 0 $.
The dilaton {\it vev} is 
\be 
s \approx \frac{\sigma}{\delta^2}\, ,
\ee
so that we require negative $\sigma$. 
The potential in this case is shown in
fig.2 where we have eliminated $\tau$ and $\phi_n$
and show the dependence on $s$ and $x$ for $K_{M\overline{M}}=e^{-60 x^2}$.
In particular (as an aside) note that there is
no barrier between large dilaton values and the minimum. This contrasts with
the multiple gaugino
condensate scenario, which has a barrier and hence an `initial value' problem 
for the dilaton. Finally, note that the imaginary axionic components of $S$, $T$ and 
$M$ are not fixed by the above but will be fixed separately by the Peccei-Quinn mechanism. 

\end{itemize} 

\noindent We repeat that the remaining fields ($S,~T,~\phi_0$) are all fixed as long  
as $x$ is fixed by one of the above mechanisms. The virtue of this set-up is therefore that 
the problem of dilaton/moduli stabilization is 
reduced to stabilizing one of the blowing up modes at a non-zero value.
The $m$ dependence in the K\"ahler potential which brings this about is unknown, 
however the set of dynamical relations between the auxilliary fields
is already very restrictive and, 
as we shall see in section 5, 
hints strongly at a solution to the SUSY flavour and CP problems.

\begin{figure}[t*]
\begin{center}
\hspace*{-0.6in}
\epsfxsize=6.55in
\epsffile{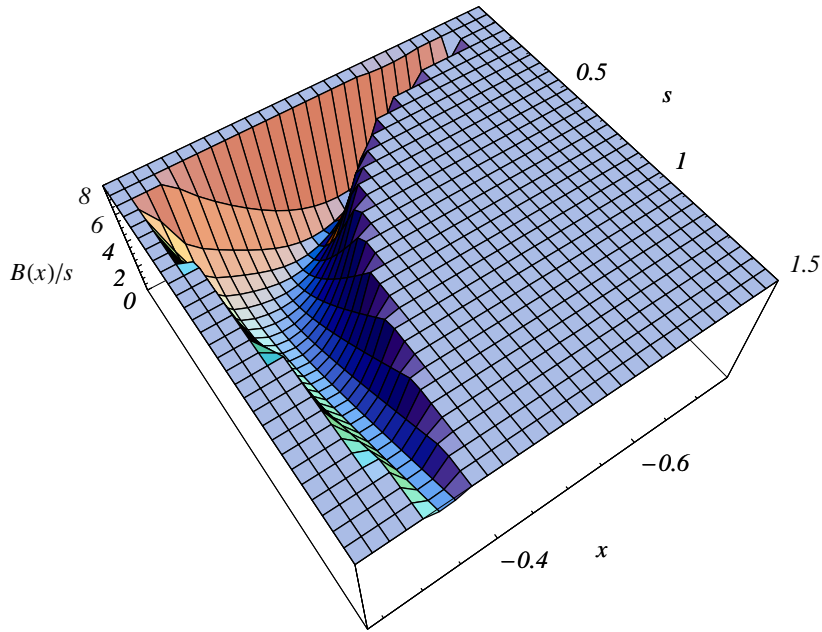}
\vspace{-3.9in}
\caption{\it The $F$-part of the potential 
(\ref{vf}) for the modular invariant case in the plane ($s,x$),
where $s=S+\bS$ and 
$x=\frac{\partial{\hat K}}{\partial M}\sim m$. The local minimum has zero 
cosmological constant. The global supersymmetric minimum is now located outside the diagram towards 
its top left (it was on the top right of the previous graph).}
\label{v2}
\end{center}
\end{figure}

\subsection{Stabilization with modular invariance}

\noindent The stabilization above is quite appealing, 
but an important aspect is that 
$W_p$ necessarily breaks any modular invariance if $\vev{W_p}$ is 
to be non-zero. One might therefore wonder how general this mechanism 
is, and in particular, if it can 
work in models that retain some or all of the 
initial modular invariance. 
In this subsection we shall show that this is indeed the case. 
In such models, the superpotential is necessarily very
different from that in eq.(\ref{toosimple}), however the same stabilization 
mechanism can be employed.

In order to include modular invariance, let us return to the superpotential,
which is now required to have the correct weight. We will consider the case of invariance 
under overall modular transformations (involving $T$), for which 
the non-perturbative contribution $W_{np}$ 
in eq.(\ref{fix2aa}) has to have weight -3. This is the case if the GS terms obey
\be
\label{modinvcond}
\delta \sigma = -2\beta \, .
\ee
In addition $U(1)_X$ invariance of $W_{np}$ requires
\be
\label{modinvcond2}
\delta_{GS} \sigma = \frac{q_0}{2\pi^2}\, .
\ee
As we saw, both of these relations are obeyed in $Z_3$ and $Z_7$ orientifold models.

To construct the rest of the superpotential,
we begin by forming gauge and modular invariant 
(and dimensionless) combinations of fields using the 
appropriate power of $W_{np}$. These we shall denote $\tilde{X}_m$;
\be 
\label{tigger}
\tilde{X}_m = \frac{X_m}{M_s^{2(1 - q_0/q_m)}} \left(\frac{W_{np}}{M_s^3}\right)^{({2q_0 - 2 q_m})/{3 q_m}}\, .
\ee
We can now write the most general expression for the superpotential as 
\be 
\label{super0}
W = W_{np} \times f(\tilde{X}_m) \, ,
\ee
where $f$ is any function of the invariants. The most trivial possibility 
with only one invariant, $\tilde{X}$, is $f=1+\tilde{X}$ in which 
case the superpotential is again just the sum of a nonperturbative and 
perturbative part
\be 
\label{super}
W = W_p + W_{np}.
\ee
(We can of course express any perturbative contribution in terms of 
$\tilde{X}_m$;
for example $\phi_2 \phi_0^2 + \phi_2 \phi^2_1 = W_{np} (\tilde{X}_2 + 
\tilde{X}_2 / \tilde{X}_1)$ and so on.)  
However in what follows, and in particular in order to spontaneously 
break CP, we will leave the expressions in the general form
of eq.(\ref{super0}). 

The imposition of modular invariance has removed one of 
our degrees of freedom since {\em a priori} eq.(\ref{super0}) gives
\be
\label{volleyball}
\frac{W_S}{W} = -\frac{8\pi^2}{N_c-1} \left(
1+ \sum_{m=1} \frac{2}{3}\frac{\tilde{X}_m f_m}{f}
\left(\frac{q_0}{q_m}-1\right)
\right)\, .
\ee
Therefore, the simplest case we can consider now has 
$\phi_0$ plus one other field (which we
shall take to be $\phi_1$) getting large {\it vevs}. 
If this is the case, we can eliminate $\hat{\phi_0}$ using the 
$D$-term constraint, and then minimize in $s$, $\hat{\phi}_1$ and $\tilde{X}_m$ 
independently. These minimizations 
again relate the {\it vevs} of the auxilliary fields;
\ba 
\label{imp_relation}
{F^1}&=&\frac{1}{p}\frac{\phi_0}{\phi_1} {F^0}\nn\\
F^M&=&-\frac{F^0}{8\pi^2\sigma\phi_0}\frac{3N_c-1+2p}{1+p} \nn\\
F^S &=& 0 \nn\\
F^{m>1} &=& 0
\, ,
\ea
where $p=\pm \sqrt{q_0/q_1}$, 
upto corrections of order $\rho_{m>1}$.
In order for these relations to be consistent, the charge 
$q_1$ must have the same sign as $q_0$, and again
we must check later on that we end up with $\vev{W_S/W} > 0$.
Eqs.(\ref{imp_relation}a,b)
imply that $F^\rho \partial_\rho g(\tilde{X}_1) = 0$ for any function 
$g(\tilde{X}_1)$. As in the non-modular invariant case, this 
will give nice phenomenological properties such as reality of $A$-terms.

Note that summing the last two equations 
and using $F_S=0$ gives the 
constraint in eq.(\ref{volleyball}).
Inserting these solutions back into $V_F$, 
we now find the $x$ dependent potential to be 
\be 
\label{BB2}
B = B_0 \frac{\delta }{\sigma^2+\lambda^2}
\left(
\sigma^2 - \sigma_0^2
+\frac{2C_0 \sigma_0^2 }{x}-\frac{C_0^2\sigma_0^2}{xB_0(x)}
\right) \, ,
\ee
where now \ba 
\sigma_0^2 &=& \frac{q_0 } 
{ 32\pi^4 \delta_{GS}\delta} \frac{(3N_c-1+2p)^2}{(1+p)^2}
\nn \\
&=& \sigma^2 \frac{(3N_c-1+2p)^2}{{(1+p)^2}{(3N_c-1)}}\, .
\ea

The resulting expression for the stabilized dilaton is 
\be
\label{vevdilaton}
s(x)  =  -\frac{\sigma^2 + \lambda^2}{\sigma B_0+{\lambda^2}
C_0 } + {\cal O}(\rho_{m>1})\, ,
\ee
where we have defined 
\ba
\lambda^2 &=& \frac{1}{\hat{A}}.\frac{q_0}{32\pi^4
|\delta_{GS}x| } \frac{ (3N_c-1+ 2 p)^2}{(1+p)^2}  \nn \\
C_0 &=& \frac{24\pi^2 }{3N_c-1+2p} .
\ea 
The remaining equations
can again be written in terms of $s(x)$;
\ba
\frac{1}{\rho_1}
& = & \frac{2 (p^2-1)}{3p(8\pi^2 s +N_c-1)}-\frac{1}{p}  \nn\\
X_1 \frac{f_{X_1} }{f}
& = & \frac{3}{2(p^2-1)} \left( 1+\frac{N_c-1}{8\pi^2s}\right) \nn\\
X_{m>1} \frac{f_{X_{m}} }{f} &=& 0 \, .
\label{imp_solutions}
\ea
The minimization of this potential is rather more 
involved than that of the previous subsection, and 
we have to be careful to ensure that both $s$ and $K_{M\overline{M}}$ are 
positive, and also that $K_{M\overline{M}}(0)=1$. 
We have identified four possibilities;

\begin{itemize}
\item Minimization with a large positive cosmological constant;
this happens quite readily for arbitrary $M$ dependence in the K\"ahler potential, due 
to the $1/x$ terms in eq.(\ref{BB2}). 

\item Minimization at $s=\infty $ corresponding to $V_F=0$.

\item The no-scale option; in this case we now have to tune away the 
$1/x$ dependence in eq.(\ref{BB2}) and the cosmological constant. 
The simplest way to do this is to 
set $\sigma^2-\sigma^2_0=\sigma^2_0 (2C_0-C_0^2/B_0)/x$ and then 
work backwards to find the required function $K_{M\overline{M}}(x)$. Note that 
$\sigma_0=\sigma$ is satisfied for $\frac{q_0}{q_1}=\frac{3N_c-1}{2}$, 
however this choice leads to negative $s$.

\item A minimum in $x$ at zero cosmological constant; the simplest 
way to find these is to `perturb' away from a no-scale solution. 

\end{itemize}

\noindent An example of the 4th case is shown in fig.3 
where we plot $B/s$ with all fields eliminated except $x$ and $s$. 
In this rather simplified case (with only one meson and one additional field)
imposing a zero cosmological constant forces rather extreme choices
of parameters in order to get a minimum at small $x$; 
in the example shown we have taken $q=1/3,~\delta = 7.6,~N_c=5$ and $p=-3.4$.
(The remaining parameters $\sigma$ and $\delta_{GS}$ are fixed by 
eq.(\ref{modinvcond}) and (\ref{modinvcond2}).)
The form of the potential is similar to that in fig.2, however 
passing over the barrier and continuing to large $x$ takes us to the 
no-scale case, and unbroken supersymmetry is now found at smaller (but finite) 
value of $x$. In addition, there is now a barrier between this minimum and 
the orbifold point. 
The minimum is at $\frac{d{B_0}}{d x}\approx 0$ as previously, so that 
eq.(\ref{mincond}) still applies.

\subsection{Relation to the heterotic string}

\noindent To complete this discussion we should relate this stabilization 
picture to that in 
heterotic strings where dilaton stabilization appears to be much more difficult. 
In particular, in the heterotic case the gauge kinetic function at one loop goes like 
$f_a=k_aS-\frac{3}{8\pi^2} c_{a} \ln\eta(T)^2$ where $T$ is again the overall 
modulus, $c_{a}$ is determined by a string computation 
and $\eta(T)=e^{-\pi T/{12} } \prod_{n=1}(1-e^{-2\pi nT})$ is the Dedekind $\eta$ function.
At large values of $T$, $\eta(T)\sim e^{-{\pi T}/{12}}$
 so that $f_a \sim k_aS+\frac{3c_a}{48\pi}T$
and one might wonder why a stabilization is not possible in this case as well,
simply by replacing $M$ with $T$. 

There are two reasons. The first is the different form of the K\"ahler potential. In the present
case the leading term goes as $m^2 = (M+\overline{M})^2$ 
as opposed to $-3 \log (T+\overline{T})$ in the heterotic case. Thus 
in the scalar potential we have $K_M \sim m $ tending to stabilize $\vev{m}$ at small 
values, as opposed to the heterotic case which has $K_T \sim -\frac{1}{T+\overline{T}}$ 
tending to push $\vev{T}$ to large values. 
The type I stabilization therefore occurs only where
$\vev{m}$ is small, of order $\delta$, 
and in this region the above approximation does not hold.

The second reason is that the stabilization relies heavily on the presence 
in the potential of anomalous $U(1)$ $D$-terms. In type I strings these contain a 
Fayet-Iliopoulos term that is proportional to $m$. Thus equating $m$ in type I with 
$T+\overline{T}$ in the heterotic string, 
would require the heterotic Fayet-Iliopoulos term to go 
like $\sim (T+\overline{T})$ rather than  
$1/(S+\overline{S})$ as is actually the case.

\section{Generating CP and flavour structure in Yukawas}

In the previous section we saw how
dilaton and moduli stabilization can occur 
in type I models assuming a single condensing gauge group and 
an anomalous $U(1)$. In following sections we shall show that 
the distribution of supersymmetry breaking amongst the 
different fields is such that SUSY contributions 
to flavour and CP violating processes are naturally suppressed. 
First however, we need a working model of flavour 
and CP violation. 

The possibilities for generating
flavour structure are restricted, since our guiding principle is to determine 
{\em all} contributions to SUSY breaking, and hence all the goldstino angles.
So, we cannot simply insert an additional Froggatt-Nielsen field without 
going back to consider the additional contribution to SUSY breaking when it gets a VEV.
Our Froggatt-Nielsen fields can therefore only be the $\phi_n$,
and we will henceforth assume that it is these fields that play a role in 
generating Yukawa hierarchies by getting vacuum expectation values.

There are many ways in which the required flavour structure could arise, but 
in order to have a working example,
consider the case with only one extra 
field $\phi_1$ with charge $2 q_1 = - q_0$, so
that we have just one relevant invariant 
$
X= \phi_1 \phi_0^2\, .
$ 
In the case without modular invariance, the most general form of Yukawa coupling can be written
 \be
\label{above}
Y_{\alpha\beta\gamma}=\tilde{\phi_0}^{q_{\alpha\beta\gamma}/q_0}g_{\alpha\beta\gamma}({\tilde{X}})
\ee
where $q_{\alpha\beta\gamma}$ is the $U(1)$ charge of the Yukawa coupling, and 
$g_{\alpha\beta\gamma}$ is any function of $\tilde{X}$.
In eq.(\ref{above}), for the case without 
modular invariance, the tilde's imply multiplication by powers of 
$M_s$ to make dimensionless quantities. 

We will see in the following sections that suppression of SUSY flavour-changing 
processes depends on choosing degenerate charges. We therefore propose that 
the $U(1)$ charges of the first and second generation charges are degenerate 
since these 
generations give rise to the most restrictive flavour changing processes. 
If $|\vev{\tilde{X}}|\ll 1$, we can choose charges such that the Yukawa couplings take the form 
\be
Y_{U_{ij}} \vev{h_2^0} \approx
\left( 
\begin{array}{ccc}
m_c & m_c & 0 \nn \\
m_c & m_c & 0 \nn \\
0 & 0 & m_t 
\end{array}
\right) 
+ {\cal O} ( m_t \tilde{X} + \ldots ) \, ,
\ee
and similar for down and lepton Yukawas,
where $m_c$ and $m_t$ are functions of $\phi_0$ only.
For the canonically 
normalized top quark, we find a mass term 
\be
\hat{m}_t = 
\vev{h_2^0} {\tilde{\phi}_0^{-(q_{H_2}+q_{t_L}+q_{t_R})/q_0}}
{\sqrt{\tau^{n_{H_2}+n_{t_L}+n_{t_R}}}} \, ,
\ee
where $n_\alpha$ are the weights of the fields. 
The $D$-term constraint imposes $|\phi_0^2| \sim |\delta_{GS}x | \tau M_P^2$ 
so that we can rewrite the top mass as 
\be 
\vev{h_2^0}
\sqrt{ |\delta_{GS} x|^{ -(q_{H_2}+q_{t_L}+q_{t_R})/q_0 } }
{\sqrt{\tau^{n_{H_2}+n_{t_L}+n_{t_R}  - (q_{H_2}+q_{t_L}+q_{t_R} )/q_0 }}} 
\left(\frac{M_P}{M_s}\right)^{-(n_{H_2}+n_{t_L}+n_{t_R} )}
\ee
In order to avoid very large values (coming from a large $\vev{\tau}$)
we impose
\be
n_{H_2}+n_{t_L}+n_{t_R}  = \frac{q_{H_2}+q_{t_L}+q_{t_R}}{q_0} \, ,
\ee
and in order to have a large top mass we require $(q_{H_2}+q_{t_L}+q_{t_R})/q_0 \approx 0$.
Imposing also that $n_{H_2}+n_{c_L}+n_{c_R}  = (q_{H_2}+q_{c_L}+q_{c_R})/q_0 $,
we find that 
\be
\hat{m}_c = \hat{m}_t \sqrt{ |\delta_{GS} x|^{ (q_{t_L}+q_{t_R}-q_{c_L}-q_{c_R})/q_0 } }
\left(\frac{M_P}{M_s}\right)^{n_{t_L}-n_{c_L}+n_{t_R}-n_{c_R} }
\ee
At leading order the Yukawa couplings have 
an accidental U(2) symmetry and hence a zero
eigenvalue. The higher order terms in $\tilde{X}$ will generally break 
this symmetry and give masses to the first generation. 
Thus, we require 
\be 
|\vev{\tilde{X}}| \sim \frac{m_u}{m_t} \, .
\ee  

Such values of $|\vev{\tilde{X}}|$, and also CP violating phases 
of order 1, can quite easily be generated during the minimization. 
The $\tilde{X}$ dependent perturbative part of the superpotential can be any function of 
$\tilde{X}$ but as a rather trivial example consider the case without modular invariance, with 
\be 
W_p = (a \tilde{X} + \frac{b}{2} \tilde{X}^2 + \frac{c}{3} \tilde{X}^3 
) M_s^3 \, .
\label{Wp}
\ee
In this expression we assume that non-renormalizable interactions between the 
fields are suppressed by powers of $M_s$ so that (since 
$\tilde{X}=X/M_s^3$)
$a,b,c$ are of order one. 
Recall that our approximations required  $|\phi_1 |\ll |\phi_0| $ so 
that, as we saw above, the minimum is close to $\partial W_p/\partial \tilde{X} = 0 $. 
The $D$-term constraint imposes $|\phi_0^2| \sim |\delta_{GS}x | \tau $ and so 
provided $\tau \gg |\delta_{GS}x|^{-1}$ (which is the case) we can check
that, for the desired value of $\tilde{X}$, this approximation is good.
It is now trivial to choose couplings 
to give the desired $\vev{\tilde{X}}\sim m_u/m_t$.

The physically observable CP violation depends only on the 
phase of $\tilde{X}$. To see this, suppose for example 
that the explicit form of the Yukawa couplings is 
\be
\left(\frac{\bra \phi_0 \ket}{M_s}\right)^{n_{ij}}H_2{Q_i}U^c_j \ + \ 
\left(\frac{\bra\phi_1\ket}{M_s}\right)^{n_{kl}}H_1{Q_k}D_l^c \, .
\ee
One of the phases can always be $R$-rotated away, and the remaining phase, 
 $\theta=(\theta_1-\frac{q_1}{q_0}\theta_0)$, is nothing other than the phase of $X$. 
(In general one needs at least two Froggatt--Nielsen fields to generate CP violation.)
Therefore, generating a physically measurable
spontaneous breaking of CP in the Yukawa couplings is equivalent to giving a phase to the gauge 
invariant $X$, i.e. by choosing $b^2-4ac<0$. 

A similar situation applies in the modular invariant case. 
Explicitly, the Yukawa couplings are 
\be
Y_{\alpha\beta\gamma}=
g_{\alpha\beta\gamma}(\widetilde{X})\left(\frac{\phi_0}{M_s}\right)^{q_{\alpha\beta\gamma}/q_0} 
\left(\frac{W_{np}}{M_s^3}\right)^{-(n_{\alpha\beta\gamma}+q_{\alpha\beta\gamma}/q_0)/3}
\ee
where $n_{\alpha\beta\gamma}$ is the modular weight of $Y_{\alpha\beta\gamma}$. 
The {\it vev} of ${\tilde{X}}$ is given by the 
equation (\ref{imp_solutions}b) with the RHS being fixed by the stabilization of $s$,
 so that we get the following equation for 
${\tilde{X}}$;
\be
{\tilde{X}}\frac{\partial f}{\partial {\tilde{X}}} +\mbox{const}f=0
\ee
Now
$f$ needs only be a polynomial of second order in ${\tilde{X}}$
in order to generate a phase for ${\widetilde{X}}$.

Notice that the CP violation does not involve the 
$T$ moduli (or $S$ in the non-modular invariant case), and these fields 
do not contribute further to it. 
In this respect the only role that  moduli
play is to transmit CP violation to the soft SUSY breaking terms. 

As we stressed, this is one of many possibilities for 
generating flavour structure with degenerate $U(1)$ charges. 
An alternative is of course to rely solely on the weights of the 
fields (and hence powers of $\tau$) to generate small 
effective micro-Yukawa couplings, as suggested in ref.\cite{imr}. 
In the modular invariant case, we will see later that  
even with different $n_\alpha$ the SUSY breaking $A$-terms
can be degenerate.
\section{Scales}

Before presenting the soft-supersymmetry breaking terms, 
we should also briefly discuss the scales of 
the moduli and dilaton {\it vevs}. In particular, since 
all supersymmetry breaking is determined, we 
find quite interesting consistency conditions on the string scale.

\subsection{Scales with degenerate $T_i$ moduli}

The moduli and dilaton of the effective 4-dimansional model 
are related to the compactification radii and string couplings as 
follows
 (see ref.\cite{imr});
\bea
\label{ti}
t_i&=&T_i+\bT_i=\frac{4}{\lambda_I} (R_iM_s)^2\\
\label{s}
s&=&S+\bS=\frac{4}{\lambda_I} \prod_i(R_iM_s)^2
\eea
For the moment  
we continue to consider the degenerate case where $T_1=T_2=T_3=T$ so that
\be
s=\tau(RM_s)^4 \, ,
\label{st}
\ee
where we have used  $|\phi_0|^2/\tau\ll1$ 
(since $\delta_{GS}x\ll 1$ in the $D$-term equation) 
to write $t\sim \tau$.

Our first concern is the compatibility between our solutions for $s$ and $\tau$ 
(which fix the gauge coupling) and  realistic gauge coupling values around  $\alpha_U \sim 1/24$.
 According to 
\be
\alpha_p=(4 \pi \Re f_p)^{-1} \, ,
\label{coupling}
\ee
this translates into: 

\noindent
1) $s\sim 4$ if the Standard Model is embedded inside D9-branes.

\noindent 
2) $\tau \sim 4$ if the Standard Model is embedded inside D5-branes.

On the other hand, the relationship between the string and Planck scales,
\be
M_s^{7-p}=\frac{\alpha_p}{\sqrt{2}}R^{p-6}M_P \, ,
\ee
reads (using equations (\ref{st}) and (\ref{coupling})),
\be
\frac{M_s}{M_P}=\frac{1}{2\pi\sqrt{2}}\frac{1}{s}\left(\frac{s}{\tau}\right)^{3/4} \, ,
\label{MS}
\ee
which gives us the supersymmetry breaking scale,
\ba
\label{doubleu}
m_{3/2} &=& |W|\frac{8\pi}{\sqrt{s\tau^3} M_P^2} e^{\hat{K}/2}\nn\\
        &\approx & 64\pi^3M_s^2 |W|/M_P^4 \, .
\ea
What values for $M_s$ are favoured by our model? 
For the moment, we are ignoring the question of the relation between the string and unification scales
 but will come back to this point later.
Concentrating on the modular invariant case, we can 
make a crude initial estimate of allowed values for $M_s$, by using
$\vev{W} \sim M_s^3$ in eq.(\ref{doubleu}) 
(assuming for the moment that $\vev{f}$ in eq.(\ref{super0}) is of order one). By implementing the
phenomenological requirement $m_{3/2}\sim m_W$ we get $M_s \sim 2.10^{15}$ GeV which 
corresponds to $\tau \sim 10^4$ if
$s \sim {\cal O}(1)$. Remarkably this is close to the conventional GUT scale, and 
can be increased by the required order of magnitude  
with almost no tuning of $\vev{f}$ as we shall shortly see.

An intermediate string scale, $M_s\sim 10^{12}$ GeV,
corresponds to $\tau \sim10^8$ and requires $\vev{f}\gg 1$\, 
\footnote{We should point out that we expect our estimates to be most 
reliable when the field theory approximation is valid and extraneous string theory effects
can be integrated out; this is guaranteed if $\phi_0 < \sqrt{\tau} M_s $ or,
because of the $D$-term constraint, $M_s > |\delta_{GS}x | M_P $, implying 
a string scale that is within a few (say 4) orders of magnitude of the 
Planck scale.}. 
Thus our crude estimate requires some improvement 
in order to treat more general choices of $M_s$, and also to estimate the 
required tuning.
Inserting the full expression for $W$ into eq.(\ref{doubleu}) we find
\ba
\label{pingpong}
M_s 
&\sim& M_P \left( \frac{e^{\frac{2\pi }{(N_c-1)\alpha_9}} m_W }{64 \pi^3 \vev{f} M_P} 
\right) ^{\frac{3 (N_c-1)}{5(3N_c-1)}} 
\\
&\sim &
2.10^{15} \gev   \, , \nn
\ea
where the final approximation is valid for large values of $N_c$ and $s\sim1$. 
In order to get $M_s \sim M_{GUT} \approx 2.10^{16}$~GeV (\ie the 
conventional MSSM GUT scale), we can for example choose $N_c=3$ or $4$
and $s\sim 1/2$ with $\vev{f}= 0.01$. Thus almost no fine tuning of $\vev{f}$
is required. It is however difficult to obtain an intermediate scale string scale.
In this simplified case (with only one condensing $SU(N_c)$ meson, we need
extremely large values of $\vev{f} > 10^8 $, which will require a significant 
amount of fine-tuning. 

There is an additional fine-tuning associated with a particular choice of $M_s$. 
In order to show where it occurs, 
let us now see how a particular value of $\tau $ (\ie $M_s$) is accommodated
by adjusting the couplings in $f$. First, the {\it vevs} of $\delta_{GS} x$ 
and $s$ are fixed completely independently via eqs.(\ref{mincond}) and 
(\ref{vevdilaton}) respectively. Our choice of $\tau$, 
{\em does} constrain $\phi_0,~\phi_1$ however, which now follow directly 
from the $D$--term constraint and eq.(\ref{imp_solutions}a). 
For example, for an intermediate string scale 
with $q_0=1/2$ and $\delta_{GS} x=0.01$ we have $\phi_0 \sim 10^3$ in 
{\em Planck} units. This is alright however, since the $\tau$ dependence cancels 
in the canonically normalized field which, since $\tau$ is dominant, is 
approximately given by $\hat{\phi}_0 = \phi_0/\sqrt{\tau}$, and therefore its {\it vev} is always 
$\approx \sqrt{\delta_{GS} x /2}$. The {\it vevs} of $\tilde{X}_{m>1}$ 
and hence $\phi_{m>1}$ are now determined 
by eq.(\ref{imp_solutions}c);
they depend on the couplings in $f$, but if they are 
similar then ${\tilde{X}_{m>1}}\sim 1$ and $\phi_m^{2q_0} \sim M_s^{2(q_0-q_m)} \phi^{2q_m}_0$.
When $q_m < q_0$ we therefore naturally get $\rho_{m>1}\ll 1$ which, recall, was one of the 
assumptions that went into the derivation of eq.(\ref{imp_solutions}).
Finally we are left with eq.(\ref{imp_solutions}b) which 
over-determines the required {\it vev} of $\phi_1$ and which is clearly independent 
of the overall $\vev{f}$.
At this point we have to satisfy this equation 
by adjusting the couplings within $f$, and this is where 
we must pay the fine-tuning price for our choice of $\tau$ (or $M_s$).

\subsection{Scales with independent $T_i$ moduli}

In the previous subsection, we found that a dilaton {\it vev} of order 1
and large degenerate moduli  {\it vevs} can explain the weak scale. 
These large {\it vevs} can be fixed with only a modest adjustment of 
couplings, and this also allows us to equate the string scale with 
the conventional GUT scale
or with the intermediate scale.
However, the assumption of degenerate moduli fields 
is rather restrictive. In particular phenomenology may not
be consistent with matter fields living on a D9-brane, and 
if all the moduli are large then the gauge couplings on the 5-branes 
are all extremely weak. In addition, stabilization usually requires $\langle s \rangle \lsim 1$.
In the degenerate moduli case therefore, 
there may be no candidate gauge couplings for the Standard Model.
We can remedy this by assuming an anisotropic compactification scheme and this is the 
issue we discuss now.

The K\"ahler potential for
independent moduli, $T_i$, is
\be
K = - \ln\, s - \sum_i \ln \tau_i + \hat{K}(m_i), 
\ee
where 
\be 
\tau_i = T_i + \bT_i  - \sum_i|C^9_i|^2
\,;\ \ \, m_i = M + \bM - \delta_i \ln \tau_i\, .
\ee
As shown in the appendix, it is still possible to compute exactly the 
inverse K\"ahler metric and hence the scalar potential. The final form is 
similar to the degenerate case, with the different moduli 
contributing separately.
In general the three $C^9_{i=1,2,3}$ transform differently with 
respect to the gauge group attached to the D9-brane.
We can for example specify that
only the $C^9_1$ condenses so that the D-term,
\be
\sum_i q_i |C^9_i|^2\frac{(1+\delta_i x)}{\tau_i}=\frac{|\delta_{GS}x|}{2},
\ee
reduces to
\be
|\phi_0|^2\frac{(1+\delta x)}{\tau_1})=\frac{|\delta_{GS}x|}{2q_0}\, .
\ee
From the form of scalar potential found in the appendix,
we see that the minimization fixes $s$, $C_1^9$, $x$ and $\tau_1$ 
precisely as before. The two remaining flat directions (corresponding to the {\it vevs} of 
$\tau_2$ and $\tau_3$) must be fixed by some other part of the theory.
(We shall comment on this presently.) 

The string scale and Planck scale are related by 
\be
\label{msp}
\frac{M_s}{M_P}=\frac{1}{2\pi\sqrt{2}(s\tau_1\tau_2\tau_3)^{1/4}}\, .
\ee
The gravitino mass is given by the same expression (\ref{doubleu}) as in the degenerate case 
(with the obvious replacement of independent radii). Repeating the previous 
analysis but now using eq.(\ref{msp}) gives the following expression 
for $M_s$:
\be
\label{tuning}
M_s
\sim  
M_P\left( \frac{e^{\frac{2\pi }{(N_c-1)\alpha_9}}m_W}{64 \pi^3 \vev{f}M_P} 
\right)^{\frac{N_c -1 }{5 N_c -3}} \, .
\ee
Large $N_c$ gives $M_s = 2.10^{15} \gev$, and
for $N_c=3,4$, this expression is compatible with 
$s\sim 1/4$ and $M_s\sim M_{GUT}$ just as before.
Now however, we also have the freedom to choose
$\tau_1\sim 4$ (\ie an embedding of the Standard Model gauge group inside $D5_1$ branes)
since, according to eq.(\ref{msp}), the relation between the string scale and $\tau_1$ involves 
the product $\tau_2\tau_3$ which is not fixed by the minimization and can be tuned instead 
(for instance $\tau_2\tau_3\sim 10^{12}$) to give $M_s \sim 2.10^{16}$ GeV. 

We can also consider the complementary scheme, in which the Standard Model gauge groups come 
from the two remaining $5$-branes and therefore $\tau_2\sim\tau_3\sim4$. In this case we 
can only adjust $\tau_1 $ in order to change $M_s$. The only difference is therefore the relation 
in eq.(\ref{msp}) which requires  $\tau_1$ to 
provide the volume factor relating $M_s$ to $M_P$ by itself.
For large $N_c$ the relation is unchanged; 
\be
M_s
\sim 
M_P\left( \frac{e^{\frac{2\pi }{(N_c-1)\alpha_9}}
m_W}{64 \pi^3 \vev{f} M_P} \right) ^ {\frac{N_c - 1 }{5 N_c - 1}} 
\ee
 
To conclude, our point in this subsection has not been to make 
any accurate estimate of parameters, but to show that with reasonable assumptions 
phenomenologically realistic values of supersymmetry breaking, string scales
and couplings are possible.
We have also highlighted where it is possible to adjust parameters in order to
get the correct size of supersymmetry breaking in the visible sector and have found that 
the fine-tuning of couplings is relatively mild. 

In the following section we examine the 
supersymmetry breaking and show that the 
structure of the model prevents the CP violating phases
phases entering into the soft-supersymmetry breaking even for an 
arbitrary numbers of fields, and for the most general Yukawa couplings. 
Before we continue, we should repeat that we will make no additional assumptions,
beyond what is necessary to stabilize the dilaton, apart from the very 
general one that CP is spontaneously broken when $f$ gets a {\it vev}, 
and that these fields enter the Yukawa coupling in some way.

\section{Structure of SUSY breaking terms}
As mentioned in the previous section, our main aim is to be able to 
generate complex Yukawa couplings by spontaneously breaking CP, 
and to ensure that this does not lead to dangerous CP violation 
in soft masses and in particular in trilinear couplings between the scalars. 
As we have seen, spontaneous breaking of CP can be driven by $U(1)$ and 
SUSY breaking (and can occur at different scales), if the $U(1)$ invariants 
$X_m$ get complex {\it vevs}. Before canonical 
normalization of the fields,
the Yukawa couplings do not depend on moduli in these models 
\cite{imr} and can be written
\be
Y_{\alpha\beta\gamma}=\left(\frac{{\phi}_0}{M_s}\right)^{\frac{q_{\alpha\beta\gamma}}
{q_0}}g_{\alpha\beta\gamma}(\tilde{X}_m)\times\left\{
\begin{array}{ll}
1&\mbox{without modular invariance}\\
W_{np}^{-(n_{\alpha\beta\gamma}+q_{\alpha\beta\gamma}/q_0)/3}&\mbox{with modular invariance}
\end{array}
\right.
\ee
where $g_{\alpha\beta\gamma}$ is an arbitrary function of the $U(1)$ invariants $X_m$ and
$q_{\alpha\beta \gamma}$ is the charge of the Yukawa.
 The phase of $\phi_0$ can always be rotated away. However, if the {\it vevs} of the $X_m$ acquire phases, they will 
induce CP violation in the CKM matrix. The minimization conditions in eq.(\ref{solutions0}b,\ref{imp_solutions}b),
can easily lead to complex 
$\langle X_m \rangle$ as we saw in section 3.
We now show that this phase in the Yukawa coupling does not feed into the 
$A$-terms and that the EDMs can therefore be suppressed.

Since we have control over the {\it vevs} of all the fields, 
it is possible to write the complete expressions for
supersymmetry breaking without having to define an arbitrary
goldstino angle. We first present the results in the case of 
an isotropic compactification for convenience. We have just concluded that
such situation  does not lead to
realistic predictions however the analysis will appear 
to be very similar in the (physically relevant) anisotropic case.

The supersymmetry breaking effects are carried 
by the auxiliary fields $F^\alpha$ which satisfy the dynamical relations 
given by (\ref{imp_relation0}) and (\ref{imp_relation}). In addition, there is the contribution of 
\be
F^T=f_T(\tau,x)+\frac{\delta\tau}{(3+\delta x)\hat{A}}F^M+\overline{\phi}_nF^n  \; \; \; \mbox{with} \; \; \;
 f_T(\tau,x)=-\frac{e^{G/2}\tau}{3+\delta x}(3+\delta B_0)
\ee
To include the possibility of fields in 9-branes or 5-branes, we will now use the general K\"ahler potential 
(\ref{completeK}) (with degenerate $T_i$ ) since visible fields may correspond to any of the four types
of $C$ fields.
As usual, we expand the K\"ahler potential around $C_\alpha=0$
in a basis in which the K\"ahler metric is diagonal in the matter fields; 
\be
K = K_0 + 
{\cal Z}_{\alpha} |C_{\alpha}|^2 + \ldots
\ee
We have:
\be
{\cal Z}_{\alpha}=\left\{
\begin{array}{lll}
s^{-1}&:&C_i^{5_i}\\
\tau^{-1}(3+\delta x)&:&C_i^9 \, C^{95_i}\\
1/2\tau^{-1/2}s^{-1/2}&:&C^{5_i5_j} \\
1/2\tau^{-1}&:&C^{95_i} 
\end{array}
\right.
\ee
Expressions for  gaugino masses, scalar masses (where $m_{3/2}=e^{G/2}$) and 
$A$-terms for a Yukawa coupling $C_\alpha C_\beta C_\sigma$ are given respectively by 
\bea 
M_a &= &\frac{1}{2} \left( Re (f_a) \right)^{-1} 
F^\alpha \partial_\alpha f_a \\
m_\alpha^2 &=& m_{3/2}^2 + V_0 - 
F^{\overline{\sigma}} F^\rho \partial_{\overline{\sigma}}
\partial_\rho \ln {\cal Z}_\alpha \\
A_{\alpha\beta\gamma} 
 &= & 
F^{\rho} 
\left( K_{\rho}+ \partial_{\rho}\ln Y_{\alpha\beta\gamma}  - \partial_{\rho} 
\ln
\left( 
{\cal Z}_\alpha{\cal Z}_\beta {\cal Z}_\gamma
\right) \right) 
\eea

\vspace{0.3cm}

\noindent {\underline{\it Gaugino Masses}};\\

\vspace{-0.4cm}
For gauge groups that live on the D9-brane the gauge kinetic function is $f_a= S+\sigma_a M$ 
so that 
\be 
M_{9} \approx\frac{\sigma_9F^{M}}{s}
\ee
 These relatively small $D9$ gaugino mass terms
arise solely due to the non-zero value $F^M$ in a manner 
suggested in ref.\cite{pheno1}. 
For gauginos associated with $D5$ branes we find
\be
M_{5_i \, a} 
\approx  
\frac{F^{T_i}}{\tau_i}+ \frac{\sigma_i}{\sigma_9}\frac{s}{\tau_i}M_9
\ee

\vspace{0.3cm}

\noindent {\underline{\it Mass-squareds}};\\

\vspace{-0.35cm}
Because of the {\it no scale} structure of the K\"ahler potential as well as  $F^S=0$ we have 
\be
F^{\overline{\sigma}} F^\rho \partial_{\overline{\sigma}}
\partial_\rho \ln {\cal Z}_{\alpha}=
\left|f_T+\frac{\delta\tau}{(3+\delta x)\hat{A}}\right|^2\partial_{\tau}^2\ln {\cal Z}_{\alpha}
=\frac{\alpha m_{3/2}^2}{(3+\delta x)^2}\left|3-\delta\frac{\delta \bW_M}{\bW}\right|^2
\ee
so that
\be
m^2_{\alpha}=m^2_{3/2}\left[1-\frac{\alpha}{(3+\delta x)^2}(3-\frac{\delta\sigma}{s})^2\right]+V_0
\ee
where
\be 
\alpha =-\tau \partial_T\ln{\cal Z}_{\alpha}= \left( 1 \, , \, \frac{1}{2}\, ,0 \, \right)
\,\,\, \mbox{ for } \,\,\, \left(
C_i^9 \, C^{95_i} \, C^{5_i}_{j\neq i} \,\, , \,\, 
 C^{5_i5_j} \,\, , \,\, C_i^{5_i} \ \right)\,\, \ \mbox{respectively}.
\ee
Generally, $\delta\sigma \ll 1$ leading to
\be
m^2_{\alpha}\approx  m_{3/2}^2 (1-\alpha) +V_0\, .
\ee
Note that in the case without modular invariance we found $V_0 \approx -\delta^2$
leading to 
$m^2_{\alpha=1} \approx -\delta^2$. However this tachyonic mass-squared is cancelled by 
additional $D$-term contributions as we shall shortly see.

\vspace{0.3cm}

\noindent {\underline{\it $A$-terms}}; \\

\vspace{-0.4cm}
The general expression is found to be
\be
\label{aterms}
A_{\alpha\beta\gamma} 
= 3m_{3/2} + x F^M-\frac{m_{3/2}}{3+\delta x}\left(3-\frac{\delta\sigma}{s}\right)
\left( \alpha +\beta+\gamma \right) +
F^{\rho}\partial_{\rho} \ln Y_{\alpha\beta\gamma}\, .
\ee
We will concentrate on the last term in (\ref{aterms})
since at first sight there seems to be a danger that it strongly violates CP.
However, it is important to take account of all contributions here because,  
although $F^m \ll F^0$, the $F^m\partial_m$ are of the same order as   
$F^0\partial_0$. Once we include all terms, a cancellation takes place.
Indeed it is clear from eq.(\ref{imp_relation0}a) and (\ref{imp_relation}a,b) that the final 
piece of the $A$-term simply counts the $U(1)$ charge of the Yukawa coupling since
\be
F^{\rho}\partial_{\rho} \ln Y_{\alpha\beta\gamma} 
 =
\frac{F^0}{q_0 \phi_0} 
\sum_{n} q_n \phi_n \partial_n \ln Y_{\alpha\beta\gamma}=\frac{F^0}{\phi_0} \frac{q_{\alpha\beta\gamma}}{q_0}
\ee
and in the modular invariant case
\be
F^{\rho}\partial_{\rho} \ln Y_{\alpha\beta\gamma} 
 =
\frac{F^0}{\phi_0} \frac{1}{1+p}\left[ p\frac{q_{\alpha\beta\gamma}}{q_0}-\alpha-\beta-\gamma\right] \, .
\ee
The $A$-terms are automatically real in both cases. Note that in the modular
invariant case there is a choice of parameters for which the dependence on 
the weights of the fields cancels. Thus, as we mentioned above, 
flavour hierarchies could be generated entirely by the weights, with 
$U(1)$ charges and hence $A$-terms remaining completely degenerate.

\vspace{0.3cm}

\noindent{\underline{\it D-term contributions}};\\

\vspace{-0.4cm}
Finally we need to consider the 
$D$-term {\it vevs}, which do not vanish precisely but generally get a {\it vev} of 
order $m_W^2$. Indeed we can develop the potential in $\hat{\phi}_0$ 
around the minimum;
\be 
V = \frac{g^2}{2}D^2 + m_{\phi_0}^2 |\hat{\phi}_0 |^2 +\ldots
\ee
where the hat stands for the canonically normalized field, 
$\hat{\phi_0}=\phi_0 \frac{3+\delta x}{\tau}$. 
Minimizing in $\hat{\phi_0}$ gives 
\be 
\label{diff}
\vev{D} = - \frac{  m_{\phi_0}^2 }{q_0 g^2} \, ,
\ee
Thus, although the approximation $\vev{D}=0$ is very accurate 
for determining the {\it vevs} of fields (\ie $\hat{\phi_0}$ is only shifted by 
${\cal{O}}(m_W) $ from the naive $\vev{D}=0$ value), eq.(\ref{diff}) gives an
additional degenerate contribution to the mass squareds of
\be 
\Delta m^2_\alpha = g^2 q_\alpha \vev{D}=\frac{q_\alpha}{q_0}m_{\phi_0}^2\, .
\ee
This is precisely the sort of contribution that the anomalous $U(1)$ mediation 
idea hopes to take advantage of in order to solve the SUSY flavour problem. 
In the present case however, the contribution is likely to be small because
the $F$-term contribution to the mass-squared of $\phi_0$ is given by 
$m_{\phi_0}^2\approx -\delta^2$, and therefore
$\Delta m_\alpha^2 \approx \delta^2 $. Thus the $D$-term 
contribution is as small as other degenerate contributions that we 
have, upto this point, been neglecting. 

If the assignment of fields is such that the 
meson field had a different modular weight, 
$\alpha$, then the $D$-term contribution could be more important.
It seems likely therefore that there 
exist cases (like $\phi_0$ not corresponding to $C^9_i$) in which a significant 
proportion of the supersymmetry breaking is mediated 
by the anomalous $U(1))$.\\

\bigskip 

To summarize, the interesting feature we find is 
that the complex $X_m$-dependent pieces cancel at zeroth order 
in $|{\phi_{m>1}}/{\phi_0}|^2$ so that  the phase of 
$A_{\alpha\beta\gamma}$  is naturally suppressed by powers of $|{\phi_{m>1}}/{\phi_0}|^2$.
This result 
applies for any number of $\phi$ fields provided that one of them dominates the D-term 
(two of them in the modular invariant case).
Indeed, the structure of $A_{\alpha\beta\gamma}$  does not depend on how many other 
condensing matter fields there are, what their charge is or what their Yukawa couplings are.
Therefore, degeneracy and reality of $A$ terms is a general feature of our model with the main
assumption being that one condensate dominates the $D$-term in the non-modular invariant 
case (and two condensates in the invariant case). It is this assumption that
forces the minimization condition to be $W_{\tilde{X}_m} \sim 0$ leading to the 
dynamical $F$-term relations in (\ref{imp_relation0}a).

We now turn to the anisotropic compactifications.
The structure of soft terms can still be predicted if we assume that the 
stabilization of the $\tau_2$ and $\tau_3$ moduli happens at a 
lower scale. Note that there is no reason for {\em all} the moduli 
(or indeed any other hidden fields) to be equally involved in 
generating and communicating the soft terms, and in breaking CP. 
If the condensing mesons couple only to $\tau_1$ as we assume here, 
then $\tau_2$ and $\tau_3$ do not need to play any role in
these processes, and if their stabilization happens
at a lower scale their effect will be negligible.
As a specific example, if these fields are fixed by gaugino condensation 
taking place on the $D5_2$ and 
$D5_3$-branes (possibly in conjunction with additional anomalous $U(1)$'s), 
then the relevant nonperturbative $T$-dependent contributions in the superpotential
must obey $e^{-4\pi^2\tau_{i}}\ll e^{-4\pi^2s}$ since phenomenology requires
$\tau_2\sim \tau_3 \gg 1$. Therefore
phenomenology dictates that any $T$ dependence in the superpotential be exponentially suppressed
so it is consistent to treat the stabilization of $S$, $\tau_1$ and $\phi_n$ separately from that 
of the remaining moduli. 

Under this general assumption, the most general expressions for $F$-terms,
including arbitrary anomalous $U(1)$, are as follows;
\bea
\label{newform2}
F^S&=&0\nn\\
&&\nn\\
F^{T_1}&=&\overline{\phi_0} F^0 -\frac{e^{G/2} \tau_1}{1+\Delta^1.x}\left[ 
1-\bphi_m \frac{\bW}{\bW}-\frac{\Delta^1\sigma}{s}\right] \nn\\
&&\nn\\
F^{T_{j=2,3}}&=&  -\frac{e^{G/2} \tau_j}{1+\Delta^j.x} 
\left(1-\frac{\Delta^1.\sigma}{s} \right)\nn\\
&&\nn\\
F^{M_k}&=& \left\{
\begin{array}{ll}
-\frac{F^0}{4\pi^2 \sigma_k \phi_0 } & \hspace{1cm}\mbox{{\footnotesize{without mod. inv.}}} \nn\\
& \nn\\
-\frac{F^0}{8\pi^2 \sigma_k \phi_0 } 
\frac{(3N_c-1+2p)}{(1+p)} & \hspace{1cm} \mbox{{\footnotesize{with mod. inv.}}} \nn 
\end{array}
\right. \nn\\
&&\nn\\
&&\nn\\
F^{m_1}&=& 
\left\{
\begin{array}{ll} 
\frac{q_m \phi_m}{q_0 \phi_0} F^0 & \hspace{3cm}\mbox{{\footnotesize{without mod. inv.}}} \nn\\
& \nn\\
\frac{\phi_m}{p\phi_0} F^0 &  \hspace{3cm} \mbox{{\footnotesize{with mod. inv.}}}
\nn 
\end{array}
\right. \nn\\
&&\nn\\
F^{m_{j=2,3}}&=& 0
\, ,
\eea
where $m_i$ labels fields coupling to $\tau_i$ in the K\"ahler potential, 
and where $\Delta_k^i$ is as defined in the appendix.
Note that each $F^{T_i}$ contributes equally to the supersymmetry breaking
even though the $\tau_i$ may be very different.
Inserting these expressions into the equations for the soft terms in the simpler 
non-modular invariant case gives
\ba
m^2_\alpha 
& = &  m_{3/2}^2 (1-\alpha)\, \nn\\
A_{\alpha\beta\gamma} 
&=& m_{3/2} 
\left(1-\frac{\delta\sigma}{s}\right)
\left[ 1 -(\alpha +\beta +\gamma) -
\frac{1}{2\pi^2s |\delta_{GS}x|}(q_{\alpha\beta \gamma} +
{\cal O}\left|\frac{\phi_m}{\phi_0}\right|^2)\right]\, ,
\ea
where $\alpha$, $\beta$, $\gamma$ are now the sum of the weights (\eg $\alpha= \sum_i \alpha_i$), 
and where here we only show the $A$-term for Yukawa couplings between fields
that couple to the same $\tau_i$. Again we see that the soft terms can be 
degenerate and that the $A$-terms are real.

\section{Further phenomenological issues}

In the previous sections we proposed a mechanism for spontaneously 
breaking CP in effective type I models with a stabilized dilaton.
We found that, if there is no $\tilde{X}$ dependant 
terms in the K\"ahler potential, then it predicts real supersymmetry breaking 
terms irrespective of the superpotential. In addition, the susy breaking is controlled by 
$U(1)$ charges, so that a particular choice can give universality in the 
susy breaking as well. 
This looks like a good start for solving the susy CP and flavour problems, however
there are a number of issues which remain. In particular, $\tilde{X}$ dependant 
terms in the K\"ahler potential might spoil these nice properties, and in 
addition we have still to control CP violation in the $\mu$ and $B$ terms. 

Therefore, we cannot yet claim to have a complete solution to the CP 
and flavour problems. But, guided by these aspects and 
by the need to preserve the suppression 
of CP violation and FCNCs, we will in this section reconsider some of 
the general ideas outlined in the 
introduction, and determine which of them (if any) can be implemented
in this framework.

\subsection{The generation of $\mu$ and $B$ and approximate CP}

Phenomenologically viable supersymmetric models require a 
higgs coupling $W_\mu = \mu H_1 H_2 $ and its 
corresponding soft-breaking term $V_{B} = \mu B h_1 h_2 + h.c.$, 
with $\mu \sim 1$TeV.
Generating a $\mu$-term of the right scale is an important problem
in supersymmetry, but it is likely to be especially difficult in
any `large dimension' model. 
Moreover the $\mu$-term is central to the susy CP problem since
electric dipole moments often constrain 
the phase of $\mu $ even more strongly than those of the $A$-terms~\cite{mssmedm}.
This is because
the magnitude of the $\mu$-term is dominant in the
mSUGRA models that are most frequently considered.
(It is customary to rotate away the 
phase of $\mu B$ since it appears in the higgs potential.)
There are a number of ways to generate $\mu$-terms and here we shall 
briefly recap them, and outline what they imply for our would-be 
solution to the flavour and CP problems. (See ref.\cite{polonsky}
for a full review.)\\

\vspace{0.1cm}

\noindent\underline{\em Non-renormalizable terms and the Giudice-Masiero mechanism}\\

\noindent One possibility for generating the $\mu$-term is to add 
non-renormalizable terms~\cite{casas}. In conventional supergravity,
where we have $\vev{|W|} \sim m_{3/2}M_P^2$, we can simply add 
the term $|W| H_1 H_2/M_P^2 $ to the superpotential which 
guarantees a $\mu$-term of the right order~\cite{casas,nilles}.
This term is equivalent at leading order to adding $H_1 H_2 + \hc$ to the 
K\"ahler potential  
as can be seen by making a K\"ahler transformation in 
the supergravity theory to transform one into the other; $K\rightarrow K + 
M_P^2(F + \overline{F})$ , $W\rightarrow W e^{F}$ where $F= 
H_1 H_2 /M_P^2  $ ~\cite{gm,nilles}. Adding a $H_1H_2$ term in the K\"ahler 
potential is the Giudice-Masieron mechanism, and such terms appear
in some heterotic string compactifications. The nett result in either case
is an additional mass term for the Higgs of order $m_W$.

In the present case (and in any model with large volume factors)
we have to be careful about scales, and also about canonically normalizing the 
higgs fields. Indeed, if we add an additional term 
\be 
W_\mu = \mu H_1 H_2 
\ee
to the supotential, then the contribution to the mass squared of the canonically normalized 
higgs fields is 
\be 
m_\mu^2 = m_{3/2}^2  \frac{\mu^2 M_P^4}{|W|^2} 
\Pi_i \vev{T_i+\overline{T}_i}^{\alpha_i+\beta_i} ,
\ee
where $\alpha_i$, $\beta_i$ are the weights of the two higgses with respect to 
$T_i$ (which can be 0, $\frac{1}{2}$, 1). 
Requiring that the $\mu$-term eventually generates a
Higgs mass term of order $m_W$ determines the magnitude of $\mu$;
\be
\label{requiredmu}
\mu = \frac{|W|}{M_P^2 \Pi_i \vev{T_i+\overline{T}_i}^{\frac{\alpha_i+\beta_i}{2}} } \, .
\ee
The notation is a little sloppy but hopefully clear; the 
$T_i+\overline{T}_i$ factors coming from the 
K\"ahler metric and the canonical normalization are the 
eventual $T_i$ {\it vevs} (not the fields themselves).
Notice that, even if the normalization factors 
$\vev{T_i + \overline{T}_i}^{\alpha_i+\beta_i}$ 
are of order one, the mass scale required for $\mu$ is very different from the 
usual $\mu\sim m_W$. For example, if $M_s\sim 10^{16} \gev$ then we would 
require $\mu \sim 10^{10}$\gev. 

The equivalent term in the K\"ahler potential is found by performing the 
K\"ahler transformation $K\rightarrow K + 
M_P^2 (F +  \overline{F})$; $W\rightarrow W e^{-F}$ with $F= 
\mu H_1 H_2 /|W| $ giving the term 
\be 
K_\mu = \hat{H}_1 \hat{H}_2 \, .
\ee
This term is the canonically normalized Giudice-Masiero term and 
adding a term like this to the K\"ahler potential seems to be
the most attractive way to guarantee higgs mass terms of the right order. 

Despite this we must still pay attention to possible non-renormalizable terms in 
the superpotential because as we have seen, 
the required $\mu$ term is proportional to $ |W| \sim M_s^3 $,
whereas in general we would expect the leading terms to be $\mu \propto M_s$,
and therefore much larger. In general therefore, we have to prevent lower order
terms from  contributing 
significantly. The relevant effective non-renormalizable terms
become important at the string scale and one does not expect additional 
volume factors to appear in the higher order diagrams. The most general 
expressions for them are therefore of the form
\be 
W_{\mu} = g_W(\tilde{X}) \tilde{\phi}_0^{-\frac{(q_{H_1}+q_{H_2})}{q_0}}
\, M_s H_1 H_2 \, ,
\ee
where $g_{W}(\tilde{X}) $ is an arbitrary function with a {\it vev} of ${\cal O}(1)$. 
Since $ \phi_0 \sim \sqrt{\tau_1|\delta_{GS} x|} M_P$, these $\mu$-terms will be many orders of 
magnitude larger than the required value in eq.(\ref{requiredmu}). 
The simplest way to forbid these terms is to choose 
$ (q_{H_1}+q_{H_2})/q_0 \notin \, 0\, \cup {\cal Z_-} $\,, so that there are no 
invariant operators with positive integer powers of $\phi_0$.

Remaining contributions to $\mu$ are then suppressed by powers of 
$\vev{\phi_1}$ and may themselves lead to a phenomenologically
desirable $\mu$. Consider, for example, 
\[
X=\phi^2_0 \phi^{\frac{-2 q_0}{q_1}}_1 
\] 
with $ (q_{H_1}+q_{H_2})/q_0 > 0$. 
In this case the leading contribution to $\mu$ is
\be 
g_W(\tilde{X}) \tilde{\phi}_1^{-\frac{(q_{H_1}+q_{H_2})}{q_1}}
\, M_s H_1 H_2 \, .
\ee
We have already seen that $\tilde{\phi}_1 \ll 1$ to get the correct value of $m_{3/2}$, and 
it is possible to choose charges so that the value of $\mu$ 
in eq(\ref{requiredmu}) results.
For example, if the higgs fields couple to the same $T_i$ 
as the mesons, then 
$X\sim M_s^3$ and $\phi_0 \sim \sqrt{ |\delta_{GS} x|/\alpha_X } M_P$ 
so that $\tilde{\phi}_1 |\delta_{GS x}| \sim \alpha_X M_s^2/M_P^2  $, and 
the generated $\mu$ is 
\be
\mu_{eff} = M_s \left( \frac{M_P^2 |\delta_{GS}x|}{M_s^2 \alpha_X}
\right)^{-\frac{q_{H_1}+q_{H_2}}{2 q_0}}.
\ee
This should be compared with the required $\mu$ in eq.(\ref{requiredmu}) which
we can write as
\be
\mu = \frac{|W| \alpha_X}{M_P^2 } \, \sim \, M_s |\delta_{GS} x| \, . 
\frac{M_s^2 \alpha_X }{M_P^2 |\delta_{GS} x|}.
\ee
The two are of the same order when 
\be 
q_{H_1}+q_{H_2}-2q_0=0 \, ,
\ee
and thus for any combination of charges satisfying this condition
we can expect a $\mu$-term of the correct order to be generated by 
non-renormalizable terms. If this combination of charges 
is negative then the $\mu$ term 
will again be too large, if it is positive then we must rely on the Giudice-Masiero 
mechanism for generating the correct term.

Once we have generated a $\mu$-term of the correct order, we
automatically have a supersymmetry 
breaking $B$-term of order $m_{3/2}$, and it is simple to show that the phase of this 
term vanishes in the same way as it does for the $A$-terms. The phase of $\mu$ 
must be small even if the phase of $\tilde{X}$ is maximal because our model favours 
$|\tilde{X}| \sim m_u/m_t$ so that 
$g_W(\tilde{X})\sim 1+d\tilde{X}$ at leading order.\\

\vspace{0.1cm}

\noindent\underline{\em An additional singlet}\\

\noindent We should briefly mention the second idea for generating an effective $\mu$ term.
It is often referred to as the Next-to-MSSM (NMSSM), and relies on
an additional gauge singlet which aquires a {\it vev} of ${\cal O}(\,1\tev)$
thanks to the higgs superpotential~\cite{nmssm},
\be 
W_S= \kappa S^3 + \lambda S H_1 H_2 \, .
\ee
The resulting phenomenology is similar to that of the usual MSSM~\cite{nmssm}.
In this original version of the NMSSM, the 
superpotential has a global $Z_3$ symmetry (under which the fields are 
all rotated by a phase $e^{i\pi/3}$). This protects the 
singlet against destabilizing divergences or non-renormalizable terms 
which would otherwise generally drive it to $\vev{S} \sim M_s$~\cite{bpr,saa}. 

The class of models under investigation clearly includes the NMSSM. 
One of the invariants $X_m$ can play the role of the NMSSM singlet, $S$. 
Again a $Z_3$ symmetry can prevent the {\it vev} of this singlet being driven to $M_s$ 
by destabilizing divergences or non-renormalizable terms, and again 
one expects that when $Z_3 $ symmetry 
is broken by the electroweak phase transition, the 
invariant acquires a {\it vev} of order $1$\tev.
The analysis of the $A$-terms is unchanged, and there is no further 
contribution to CP 
violation, so that the EDMs are protected. From this point of view additional 
singlets with {\it vevs} protected by discrete symmetries are beneficial
and seem natural, however these models have other serious
difficulties. The $Z_3$ symmetry implies that the breaking of
electroweak symmetry gives cosmological 
domain walls with a typical mass per unit area of $m_W^3$~\cite{walls}. 
One possible way to avoid destabilizing divergences 
and yet break the $Z_3$ symmetry in the global theory is to impose instead 
an $R$-symmetry in the model~\cite{saa}. There has been recent interest 
in this idea although we will not pursue it here~\cite{rsymm}. 
(For other aspects of these models and other ideas 
see ref.\cite{polonsky} and references therein.)

\subsection{Higher order corrections in the K\"ahler potential}

We now turn to the question of higher order corrections which may 
destroy the attractive properties of the soft breaking terms
that we have found at leading order. 
As we have seen, these properties are independent of the 
form of the superpotential to all orders. 
Therefore, the most dangerous terms are those coming from higher 
order contributions in the K\"ahler potential. 
For example, the K\"ahler potential can take the form 
\be 
\label{danger}
K_i^j = (\delta^i_j + K^{(1)\, i}_j(X,X^+))
 \hat{\phi}_i \hat{\phi}^j + \hc 
\ee
where $K^{(1)\, i}_j$ are some functions of the invariants. 
These higher order terms in eq.(\ref{danger}) are generally 
flavour changing and, because of the phase of 
$\vev{X}$, they will also contribute to CP violating observables such as
the $\varepsilon$ and $\varepsilon'$ parameters of the kaon system. 
We can estimate the constraint on these terms from the fact that
the additional contribution to the K\"ahler potential will cause 
a mass splitting in the squark mass-squareds of order, 
\be 
\delta^f_{ij} = \frac{m_i^2 - m_j^2}{m_i^2+m_j^2} \approx K^{(1)\, i}_j \, ,
\ee
where $f$ is the squark label.
The bounds on these parameters have been widely studied 
in the literature, and they are particularly strong for the 
1st and 2nd generation mixing, and for the combination
$\delta_{12}^f = \sqrt{(\delta^f_{LL})_{12}(\delta^f_{RR})_{12}}$. 
From $\Delta M_K $ and $\varepsilon_K$ one finds that\cite{gabbianimasiero}
\be 
Re [\delta^f_{12} ] \leqsim 6.10^{-3} \left(\frac{m_{\tilde{d}}}{1\tev}\right)
\,\,\,\, , \,\,\,\, 
Im [\delta^f_{12} ] \leqsim 5.10^{-4} \left(\frac{m_{\tilde{d}}}{1\tev}\right) \, .
\ee
The most direct way to satisfy these constraints is to tune $\vev{|X|}$ to be 
small. As we discussed earlier this means tuning the parameters in the perturbative superpotential and 
will still be compatible with setting $M_s = M_{GUT}$ if we choose the rank of the 
condensing hidden gauge group, $N_c$, correctly. 

More dangerous higher order corrections come from terms of the form
\be
\label{danger2}
{\cal C} \frac{|\hat{\phi}_0|^2}{M_s^2} \hat{\phi}_i \hat{\phi}^j \sim 
{\cal C}|\delta_{GS}x|\frac{M_P^2}{M_s^2}\hat{\phi}_i \hat{\phi}^j 
\ee
where ${\cal C}$ represents some coupling factor between $|\hat{\phi}_0|^2$ and the $\hat{\phi}_i$'s.
To forbid these requires ${\cal C}\ll \frac{{M_s^2}/M_P^2}{|\delta_{GS}x|}\sim 10^{-6}/{|\delta_{GS}x|}$.
We reasonably expect $|\delta_{GS}x|\sim 10^{-3}$ so that the constraint becomes ${\cal C}\ll 10^{-3}$. 
We recall that $\phi_0$ corresponds either to $C^9_i$ or $C^{5_k}_j$ fields. Also, tree level interactions 
between $C$ fields are described by the superpotential \cite{berkooz,imr}
\bea
\label{superpot9}
W_9 & =&   {C_1^{9}} {C_2^{9}} {C_3^{9}} \ + \   {C^{5_1 5_2}}  {C^{5_3 5_1}}  {C^{5_2 5_3}} \ + \ 
\sum_{i=1}^3  {C_i^{9}}  {C^{9 5_i}}   {C^{9 5_i}} \\
\label{superpot5}
W_5 & = &\sum_{i=1}^3 
\left(  {C_1^{5_i}} {C_2^{5_i}} {C_3^{5_i}}   
 +  {C_i^{5_i}} {C^{9 5_i}}  {C^{9 5_i}} \right) \\
& &  + \sum_{i\neq j \neq k=1}^3 \left( {C_{j}^{5_{i}}} {C^{5_{i} 5_{k}}}{C^{5_{i} 5_{k}}}
 +  \frac{1}{2} {C^{5_{j} 5_{k}}} {C^{9 5_{j}}} {C^{9 5_{k}}} \right) 
\nn
\eea
 To suppress the interactions (\ref{danger2}), we see from (\ref{superpot9}) that visible fields should not 
be $C^9_i$ nor  $C^{95_i}$ if $\phi_0$ is associated with 
$C^9_i$. However, they can be any other $C$ fields, in which case the interactions 
(\ref{danger2}) will come from loop diagrams and we expect them to be suppressed and obey the constraint 
${\cal C}\ll 10^{-3}$. Similarly if $\phi_0$ is associated with $C^{5_k}_j$. In this case, visible fields should not be 
$C^{5_k}_j$ nor $C^{5_i5_j}$ according to (\ref{superpot5}).

\subsection{Relating CP violation and flavour changing}

Finally we remark on another scheme for addressing the susy CP problem, 
which is to associate CP violation with flavour violation. This idea
arises naturally when the Yukawa couplings are associated with adjoint fields 
that acquire {\it vevs} ~\cite{bailin} and the resulting couplings are hermitian 
in flavour space.
Such a scheme can be incorporated into the present 
framework by putting the $X$ fields into the adjoint representation 
of a horizontal flavour symmetry. 
Note that a requirement for this to work is that the supersymmetry breaking 
dynamics to not contribute further to CP violation~\cite{bailin}, as is indeed the 
case here. (The role of the $F$-term {\it vevs} is merely one of transmitting 
the CP violation to the visible sectors.)

It is clear, since we have been 
careful to maintain complete control over 
the dynamics of dilaton stabilization and over the spontaneous 
breaking of CP, that this idea goes through unchanged. In particular, 
we diagonalize the K\"ahler metric in eq.(\ref{danger}) by making 
unitary transformations. However the $A$-terms and Yukawa couplings are 
hermitian. Diagonalization of the Yukawas therefore also involves a unitary 
transformation and the $A$-terms remain hermitian with real diagonal components.
This scheme keeps CP violation out of the $\mu$-term
and the flavour diagonal $A$-terms, and  
the contribution to EDMs is automatically suppressed; 
CP violation is always, as observed in nature, associated with flavour changing.

\section{Summary}
We have studied supersymmetry breaking by a single gaugino condensation in the context of type 
IIB orientifolds. While there does not yet exist any realistic particle spectrum in the simplest
constructions \cite{imr}, our goal was to bring out some phenomenological features 
that are expected to be typical of these theories. 
We found that it is possible to stabilize the dilaton and moduli fields
at {\it vevs} which are in agreement with both realistic gauge couplings and an electroweak 
supersymmetry breaking scale.
It also predicts a string scale close to the conventional unification scale $M_{GUT}$. 
The stabilization utilizes an anomalous $U(1)$ symmetry and relies 
heavily on the presence of twisted moduli fields, $M$, 
that appear in the K\"ahler potential.
In addition to these nice properties the stabilization 
incorporates spontaneous CP breaking leading to
complex Yukawa couplings and a CP violating phase in the CKM matrix. In contrast,
soft-masses, and the $A$ and $B$-terms are guaranteed to be real at leading order,
and depend only on the choice of $U(1)$ charges, 
hinting strongly at a possible solution to the susy flavour and CP problems. 
Suppressing the higher order contributions to CP and flavour changing 
requires some constraints on the $U(1)$ charges 
and this favours some non-universality in the supersymmetry breaking.

We should emphasize the major role played by the anomalous $U(1)$ in our analysis. 
First it generates an $M$-dependent Fayet--Iliopoulos term and consequently fixes the 
{\it vev} of the meson via the $D$-term equation. Second, it plays a role in spontaneous CP
violation; the size of the phase in the CKM matrix is given by the phase of the $U(1)$
invariants. This depends on the couplings between
the non-renormalizable interactions in the perturbative superpotential.
Finally the $U(1)$ symmetry governs the flavour structure on two levels: 
not only does the $U(1)$ charge determine the $A$-terms, but also 
it is useful to produce a Froggatt--Nielsen like hierarchy in the quark 
mass matrix. The relation between charges, soft-terms and Yukawa couplings is similar to that 
of ref.\cite{carlos} and the phenomenological
consequences are expected to be the same. One particularly interesting conclusion is therefore that
 the {\it vev} of the $U(1)$ 
invariants determines both the Yukawa hierarchy and the string scale. 
Our final picture is the following:
\begin{itemize}

\item if $arg(\tilde{X})\sim {\CO}(1)$, we have maximal CP violation in the CKM matrix, 
however the soft terms are real. 
Also, the phase of $\mu$ is small because $|\tilde{X}|<1$. We can choose charges such that $A$-terms are universal 
and then FCNCs and EDMs are both suppressed at leading order. However, avoiding FCNCs at higher 
order requires some non-universality in the $A$-terms.

\item if $arg(\tilde{X})$ is small, this is a natural scenario for approximate CP~\cite{nir}. 
Now the non-universality is required to account for the observed 
value of $\varepsilon_K$ and $\varepsilon'_K$ through supersymmetric diagrams.

\end{itemize}

\noindent These two possibilities 
for solving the susy flavour and CP problems are 
already well-known general ideas in the literature. 
However, here they are the outcome of the specific dynamics of dilaton and moduli stabilization
leading to dynamical relations between the auxilliary fields.

\section*{Acknowledgements}
We are very grateful to Carlos Savoy for helpful collaboration 
and wish to thank St\'ephane Lavignac for enlightening discussions.
SAA is supported by a PPARC Opportunity Grant.

\section*{Appendix: 
Inverse K\"ahler metric and scalar potential in the case $T_1 \neq T_2 \neq T_3$}
In \cite{as}, we detailed the computation of the inverse K\"ahler metric and the scalar potential
in the case where all moduli $T_i$ are assumed to be degenerate \ie $T_1=T_2=T_3=T$. 
In this appendix we relax this assumption and expose the results in the most general case. It appears that
the potential can still be expressed in a simple form similar to the degenerate case. 
  
For independent $T_i$, the K\"ahler potential is:
\be
K = - \ln\, s - \sum_{i=1}^3 \ln \tau_i + \hat{K}(m_k), 
\ee
where 
\be 
\tau_i = T_i + \bT_i  - |\phi_i|^2
\,;\ \ \, m_k = M_k + \bM_k - \sum_i\Delta_k^i \ln \tau_i\, .
\ee
From the K\"ahler equation (\ref{completeK}), we have only kept the $C^9_i$ matter 
fields (for convenience, $C^9_i$  are denoted $\phi_i$) since only the fields
charged under the D9-brane gauge groups will condense. Note that $\phi_i = \phi_{i\, n}$ are 
vectors, so that we are allowing for the most general case.
Note that we changed the notation $\delta$ to $\Delta$ since it now carries two indices and could 
be confused with the Kronecker symbol.

We get the following for the first derivative of the K\"ahler potential:
\be 
K_\alpha = \left( 
-\frac{1}{s}
\, , \, 
- \frac{ (1+\Delta^i . x) }{\tau_i}
\, , \, 
x_k 
\, , \, 
\bphi_i  \frac{(1+\Delta^i . x) }{\tau_i} \right) \, . 
\ee
Here we have introduced $x_k = \pt \hat{K} /\pt M_k $, and 
have defined a dot product, $\Delta^i . x \equiv \sum_k\Delta^i_k x_k $.
Differentiating again we get
\be 
K_{\alpha\bbeta} = \left( 
\begin{array}{cc}
\frac{1}{s^2} & 0 \nn\\
0 & K_{a\overline b} 
\end{array} \right) .
\ee 

To express $K_{a\overline b}$ and its inverse we define 
\ba
J_{kk'} &=& \frac{\pt^2 \hat{K} }{\pt M_k \pt \bM_{k'} }\nn\\
A_{kk'} &=& (J^{-1})_{kk'}  + \sum_i\Delta_k^i \frac{1}{(1+\Delta^i.x)}\Delta_{k'}^i \nn\\ 
\CC_{ij}       &=& \delta_{ij} \frac{1+\Delta^i.x}{\tau_i^2}+ \frac{\Delta^i.J.\Delta^j}{\tau_i\tau_j} \nn\\
\Delta^i.J.\Delta^j &=&\Delta_k^i J_{kk'} \Delta_{k'}^j
\ea
and then have 
\bea
K_{a\overline b} &=& \left( 
\begin{array}{ccc}
K_{T_i\bT_j} \ \ & \ \ K_{T_i\bM_k} \ \ & \ \ K_{T_i\bphi_{j'}}\nn\\
K_{M_{k'}\bT_j} \ \ & \ \ K_{M_{k'}\bM_k} \ \ & \ \ K_{M_{k'}\bphi_{j'}}\nn\\
K_{\phi_{i'}\bT_j} \ \ & \ \ K_{\phi_{i'}\bM_k} \ \ & \ \ K_{\phi_{i'}\bphi_{j'}}
\end{array} \right) \\
&=& \left( 
\begin{array}{ccc}
\CC_{ij} \ \ & \ \ -\frac{(\Delta^i.J)_k}{\tau_i} \ \ & \ \ -\phi_{j'} \CC_{ij'} \nn\\
 - \frac{(\Delta^j.J)_{k'}}{\tau_j} \ \ & \ \ J_{k'k} \ \ & \ \ \phi_{j'}\frac{(\Delta^{j'}.J)_{k'}}{\tau_{j'}}
 \nn\\
 -\bphi_{i'} \CC_{i'j} \ \ & \ \ \bphi_{i'}\frac{(\Delta^{i'}.J)_{k}}{\tau_{i'}} \ \ & \ \ 
 \bphi_{i'}\phi_{j'} \CC_{i'j'} +\delta_{i'j'}
  \frac{1+\Delta^{i'}.x}{\tau_{i'}} 
\end{array} \right) 
\eea 

The inverse of this matrix is 
\be
K^{\balpha\beta} = \left( 
\begin{array}{cc}
s^2 & 0 \nn\\
0 & K^{{\overline a} b} 
\end{array} \right) 
\ee
 
\bea
 K^{{\overline a} b} & = &\left( 
 \begin{array}{ccc}
 K^{\bT_iT_j} \ \ & \ \ K^{\bT_iM_k} \ \ & \ \ K^{\bT_i\phi_{j'}}\nn\\
K^{\bM_{k'}T_j} \ \ & \ \ K^{\bM_{k'}M_k} \ \ & \ \ K^{\bM_{k'}\phi_{j'}}\nn\\
K^{\bphi_{i'}T_j} \ \ & \ \ K^{\bphi_{i'}M_k} \ \ & \ \ K^{\bphi_{i'}\phi_{j'}}
\end{array} \right) \\
 & = &\left( 
 \begin{array}{ccc}
 \frac{(\tau_i^2+\tau_i|\phi_i|^2)\delta_{ij}}{(1+\Delta^i.x)} \ \ & \ \ 
 \frac{\tau_j\Delta^j_{k'}}{(1+\Delta_j.x)} \ \ & \ \ \frac{\phi_i \tau_i}{(1+\Delta^i.x)}\delta_{ij'} \nn\\
 \frac{\tau_j\Delta^j_k}{(1+\Delta_j.x)} \ \ & \ \ A_{k'k} \ \ & \ \ 0 \nn\\
\frac{\bphi_{i'} \tau_{i'} }{(1+\Delta_{i'}.x)} \delta_{i'j} \ \ & \ \ 0 \ \ & \ \ 
\frac{\tau_{i'}}{(1+\Delta^{i'}.x)}  \delta_{i'j'}
\end{array} \right)  
\eea
The hidden sector group being on a D9-brane, 
the superpotential does not depend on the $T_i$. 
One can easily check that the $F$-part of the scalar potential, 
\be
V_F=e^G(-3 + G_{\alpha} K^{\alpha \bbeta} G_{\bbeta})=e^G B \ \ \ \ \mbox{where} \ \ \ \  G=K+\ln|W|^2
\ee
may be written as:
\be
V_F=e^G\left(\Delta.B_0
+s^2|G_S|^2+\sum_{i=1}^3\frac{\tau_i}{(1+\Delta^i.x)}\left|\frac{W_{\phi_i}}{W}\right|^2+
\sum_{kk'}(B_{0,k}+\frac{W_k}{W})A_{kk'}(B_{0,k'}+\frac{W_{k'}}{W})\right)
\ee
where we have defined
\be
\Delta.B_0\equiv \sum_k \Delta_k B_{0,k} , \ \ \ \ \Delta_k=\sum_i\Delta^i_k
\ee
\be
B_{0,k}=x_k-(A^{-1})_{kk'}\Delta_{k'}
\ee
Like in the degenerate case, the dilaton is fixed because $x_k$ and $B_{0,k}$ get a {\it vev}.
There is also the  $D$-term contribution,
\be
V_D=\frac{g_X^2}{2} |D_X|^2\, ,
\ee
where
\be
D_X = \sum_i q_i|\phi_i|^2 \frac{(1+\Delta^i.x)}{\tau_i}+ \delta _{GS_k} \frac{x_k}{2}
\ee
We see that if one field dominates, for example $\phi_2$, then the $D$-term 
equation $D_X=0$ does fix the ratio  $|\phi_2|^2/\tau_2$ in terms of $x$. One can also assume that $C^9_1$
and $C^9_3$ are not charged under $U(1)_X$. While our model looses 
some predictability (the string scale will remain a free parameter with $\tau_1$ and 
$\tau_3$), one advantage is that the dilaton and twisted moduli (and therefore the gauge
couplings on the $D9$-brane) are fixed
whatever the values of $\tau_1$ and $\tau_3$.


\end{document}